\documentclass[12pt]{article}
\usepackage{amsmath}
\newcommand{\blind}{0}

\makeatletter
\renewcommand\@biblabel[1]{\indent}
\makeatother

\newcommand{\vect}{\textup{vec}}

\newcommand{\diag}{\textup{diag}}
\newcommand{\avar}{\textup{avar}}

\newcommand{\vech}{\mbox{vech}}

\newcommand{\rank}{\mbox{rank}}
\newcommand{\Span}{\mbox{span}}
\newcommand{\cov}{\mbox{Cov}}
\newcommand{\tr}{\mbox{tr}}

\newcommand{\bX}{\mbox{$\mathbf{X}$}}
\newcommand{\bx}{\mbox{$\mathbf{x}$}}

\newcommand{\bbeta}{\mbox{$\boldsymbol\beta$}}
\newcommand{\balpha}{\mbox{$\boldsymbol\alpha$}}

\newcommand{\bvarepsilon}{\mbox{$\boldsymbol\varepsilon$}}

\newcommand{\bnu}{\mbox{$\boldsymbol\nu$}}
\newcommand{\bxi}{\mbox{$\boldsymbol\xi$}}
\newcommand{\bOmega}{\mbox{$\boldsymbol\Omega$}}
\newcommand{\bSigma}{\mbox{$\boldsymbol\Sigma$}}
\newcommand{\bGamma}{\mbox{$\boldsymbol\Gamma$}}

\newcommand{\bY}{\mbox{$\mathbf{Y}$}}
\newcommand{\by}{\mbox{$\mathbf{y}$}}
\newcommand{\bI}{\mbox{$\mathbf{I}$}}

\newcommand{\bA}{\mbox{$\mathbf{A}$}}
\newcommand{\bB}{\mbox{$\mathbf{B}$}}
\newcommand{\bC}{\mbox{$\mathbf{C}$}}
\newcommand{\bD}{\mbox{$\mathbf{D}$}}

\newcommand{\bP}{\mbox{$\mathbf{P}$}}
\newcommand{\bQ}{\mbox{$\mathbf{Q}$}}
\newcommand{\bR}{\mbox{$\mathbf{R}$}}

\newcommand{\bV}{\mbox{$\mathbf{V}$}}

\newcommand{\bZ}{\mbox{$\mathbf{Z}$}}

\newcommand{\bzeta}{\mbox{$\boldsymbol\zeta$}}

\newcommand{\bPhi}{\mbox{$\boldsymbol\Phi$}}

\newcommand{\R}{\mbox{$\mathbb{R}$}}

\newcommand{\OLSVAR}{\scriptsize\textup{OLSVAR}}
\newcommand{\EVAR}{\scriptsize\textup{EVAR}}
\newcommand{\RRVAR}{\textup{\scriptsize RRVAR}}
\newcommand{\REVAR}{\scriptsize\textup{REVAR}}

\usepackage{setspace}
\usepackage{float}

\newtheorem{lem}{Lemma}%[section]
\newtheorem{Theorem}{Theorem}[section]
\newtheorem{Definition}[Theorem]{Definition}

\newtheorem{proposition}{Proposition}

\usepackage{graphicx}
\usepackage{caption}
\usepackage{subcaption}
\usepackage{amsfonts}

\usepackage[font=small,skip=-6pt]{caption}

\usepackage{multirow}
 \usepackage{booktabs} % The booktabs package also provides \cmidrule that can be with l or r to shorten the line in the left or right end, respectively. 

\usepackage{enumitem} 
\makeatletter
\renewcommand*\env@matrix[1][\arraystretch]{%
  \edef\arraystretch{#1}%
  \hskip -\arraycolsep
  \let\@ifnextchar\new@ifnextchar
  \array{*\c@MaxMatrixCols c}}
\makeatother

\begin{document}

\bibliographystyle{natbib}

\def\spacingset#1{\renewcommand{\baselinestretch}%
{#1}\small\normalsize} \spacingset{1}

 \date{}

%%%%%%%%%%%%%%%%%%%%%%%%%%%%%%%%%%%%%%%%%%%%%%%%%%%%%%%%%%%%%%%%%%%%%%%%%%%%%%

\if0\blind
{
  \title{\bf Reduced-rank Envelope Vector Autoregressive Models}
  \author{ S. Yaser Samadi \thanks{Corresponding author, Email: ysamadi@siu.edu
    %The authors gratefully acknowledge \textit{please remember to list all relevant funding sources in the unblinded version}
    }\hspace{.2cm} and
    %Department of YYY, University of XXX\\
  %  and \\
    H. M. Wiranthe B. Herath \\
   School of Mathematical and Statistical Sciences, Southern Illinois University\\ Carbondale, IL\\
   Department of Information Management and Business Analytics\\ Drake University, Des Moines IA }
  \maketitle
} \fi

\if1\blind
{
  \bigskip
  \bigskip
  \bigskip
  \begin{center}
    {\LARGE\bf Reduced-rank envelope vector autoregressive model}
\end{center}
  \medskip
} \fi

\bigskip

\begin{abstract}
The standard vector autoregressive (VAR) models suffer from overparameterization which is a serious issue for high-dimensional time series data as it restricts the number of variables and lags that can be incorporated into the model. Several statistical methods,  such as the reduced-rank model for multivariate (multiple) time series (Velu et al., 1986; Reinsel and Velu, 1998;  Reinsel et al., 2022) and the Envelope VAR  model (Wang and Ding, 2018),  provide solutions for achieving dimension reduction of the parameter space of the VAR model. 
 However, these methods can be inefficient in extracting relevant information from complex data, as they fail to distinguish between relevant and irrelevant information, or they are inefficient in addressing the rank deficiency problem. 
We put together the idea of envelope models into the reduced-rank VAR model to simultaneously tackle these challenges,  and propose a new parsimonious version of the classical VAR  model called the reduced-rank envelope VAR (REVAR) model. Our proposed REVAR model incorporates the strengths of both reduced-rank VAR and envelope VAR models and leads to significant gains in efficiency and accuracy. The asymptotic properties of the proposed estimators are established under different error assumptions. Simulation studies and real data analysis are conducted to evaluate and illustrate the proposed method.
\end{abstract}

\noindent%
\vfill

Journal of Business \& Economic Statistics,   {https://doi.org/10.1080/07350015.2023.2260862} %10.1080/07350015.2023.2260862

\newpage
\spacingset{1.74} % DON'T change the spacing!

\setlength{\abovedisplayskip}{7.4pt}
\setlength{\belowdisplayskip}{7.4pt}

\section{Introduction}\label{Introduction}
%\doublespacing
 
% \vspace{-.15in}
  With the recent rapid development of  information technology,  high-dimensional time series data are routinely collected in
  various fields, like
  finance, economics, digital signal processing, neuroscience, and meteorology.   Classical vector autoregressive (VAR) models are widely used for modeling multivariate time series data due to their ability to capture dynamic relationships among variables in a system and their flexibility.  These models are discussed in many time series textbooks, including   Hamilton (1994),  L\"{u}tkepohl (2005), Tsay (2014), Box et al. (2015), Wei (2019), and others.  However, the VAR model suffers from overparameterization, particularly when the number of lags and time series increases. While there are several statistical methods for achieving dimension reduction in time series (Park and Samadi, 2014, 2020; Cubadda and Hecq, 2022b; Samadi and DeAlwis, 2023a), they can be inefficient in extracting relevant information from complex data. This is because they fail to differentiate between important and unimportant information, which can obscure the material and useful information. Envelope methods (Cook et al., 2010) use reduced subspaces to link the mean function and dispersion matrix through novel parameterizations.  By identifying and removing irrelevant information, the envelope model is based only on useful information and is therefore more efficient.
  %Various methods have been developed in the literature for multivariate time series analysis,
  The literature presents various methods for multivariate time series analysis,
  including the canonical transformation  (Box and Tiao (1977),  reduced-rank VAR models   (Velu et al., 1986;  Reinsel et al., 2022),  scalar component models  (Tiao and Tsay, 1989),   LASSO regularization of VAR models  (Shojaie and Michailidis, 2010;  Song and Bickel, 2011),   sparse VAR model based on partial spectral coherence  (Davis et al., 2016),     factor modeling  (Stock and Watson, 2005;  Forni et al., 2005;  Lam and Yao, 2012),   envelope VAR models  (Wang and Ding, 2018; Herath and Samadi, 2023a, 2023b), nonlinear VAR models (Samadi et al. 2019),   and tensor-structure modeling for VAR models  (Wang et al., 2022a;  Wang et al.,  2021),  among others.

Let $\mathbf{y}_t = (y_{1t}, \hdots, y_{qt})^{'}$ be a $q$-dimensional VAR process % vector autoregressive process
of order $p$, VAR($p$), given as %(L\"{u}tkepohl, 2005; Tsay, 2014)
\begin{equation}\label{eq:2.1}
 \mathbf{y}_t = \balpha + \bbeta_1 \mathbf{y}_{t-1}+\bbeta_2 \mathbf{y}_{t-2}+ \hdots +\bbeta_p \mathbf{y}_{t-p} + \boldsymbol\varepsilon_t,~~~ t = 1,...,T,
\end{equation}
where $\bbeta_i \in \mathbb{R}^{q \times q}$ are coefficient  matrices, and $\boldsymbol\varepsilon_t \in \mathbb{R}^{q}$ is a vector white noise process with mean $\mathbf{0}$, and covariance matrix $\bSigma$, i.e., $\boldsymbol\varepsilon_t \sim WN(\mathbf{0},\bSigma)$, and $T$ denotes the sample size. Suppose $L$ is the lag operator and let $\bbeta(L)$ be the characteristic polynomial function of the model given as $\bbeta(L) = \bI_q - \bbeta_1L - \hdots - \bbeta_pL^p$. Then, the time series in \eqref{eq:2.1} is stationary if all the roots of $\det\left(\bbeta(L)\right) = 0$ are greater than one in modulus.

Modeling high-dimensional multivariate time series is always challenging due to the dependent and high-dimensional nature of the data.  Even for moderate dimensions $q$ and $p$, performing the estimation can be difficult (De Mol et al., 2008; Carriero et al.,  2011; Koop 2013).  The number of coefficient parameters in model \eqref{eq:2.1},  $q^2p$,  can dramatically increase with the dimension $q$ and the lag order $p$.
Therefore,  to improve   estimation and  make inferences  on high-dimensional VAR models, it is necessary to restrict the parameter space 
to a reasonable and manageable number of parameters.  To this end, note that the  VAR($p$) model  in  \eqref{eq:2.1} can be rewritten as a VAR($1$)  model as
\begin{equation}\label{eq:2.2}
  \mathbf{y}_t =\balpha + \bbeta \mathbf{x}_{t} + \boldsymbol\varepsilon_t,
\end{equation}
where $\mathbf{x}_{t} \in \mathbb{R}^{qp \times 1}$ is the vector of lagged variables, i.e., $\mathbf{x}_{t} = ( \mathbf{y}_{t-1}^{'}, \mathbf{y}_{t-2}^{'}, ..., \mathbf{y}_{t-p}^{'})^{'}$, 
and $\bbeta = (\bbeta_1, \bbeta_2, ...,\bbeta_p)\in \mathbb{R}^{q \times qp}$ which encompasses the coefficient matrices of lagged variables.

To address the overparameterization issue of the VAR model, we assume that the autoregressive coefficient matrix $\boldsymbol\beta$ in model \eqref{eq:2.2} has a reduced-rank structure, similar to the standard reduced-rank regression (RRR)  model (Velu et al., 1986; Anderson, 1999,  2002; Reinsel et al., 2022).
This technique improves the accuracy of the estimation of $\bbeta$ by reducing the dimensionality of $\mathbf{y}_t$ and $\mathbf{x}_{t}$. The envelope model (Cook et al., 2010) is a new dimension reduction technique that has a different perspective to achieve efficient estimation by linking the mean function and covariance matrix and using the minimal reducing subspace of the covariance matrix. By combining these two methods,  we propose a novel dimensionality reduction model for the VAR process called the reduced-rank envelope vector autoregressive 
(REVAR) model, which extends the idea of envelopes to the reduced-rank VAR model (RRVAR). As a result, our proposed method is more efficient and parsimonious than both methods alone and outperforms both.

The reduced-rank problem first arose in the multivariate regression analysis to achieve dimension reduction by restricting the rank of the coefficient matrix. Anderson (1951) considered the 
RRR  %reduced-rank regression 
problem for fixed predictors. Then, Izenman (1975) and Reinsel and Velu (1998) studied the RRR % reduced-rank regression 
in detail. Asymptotic properties for RRR %reduced-rank regression
are discussed by Stoica and Viberg (1996), and  Anderson (1999). The  RRR % reduced-rank regression
models have also been investigated by several other authors (Reinsel and Velu, 1998; %Yuan et al. 2007; 
Negahban and Wainwright 2011; Chen et al., 2013; Basu et al., 2019; Raskutti et al., 2019).

Reduced-rank model for multivariate (multiple) time series (Velu et al. 1986; Reinsel and Velu 1998; Reinsel et al., 2022) often arises in the multivariate statistics literature when coefficient matrices have low-rank structures.
The reduced-rank VAR (RRVAR) model is considered by imposing a low-rank structure on the coefficient matrix of model \eqref{eq:2.2}, i.e., rank $(\bbeta) = d < q$. 
As a result, the number of parameters is decreased and the estimation efficiency is improved.   
The analysis of  RRVAR models has connections with some known methodologies such as principal component analysis (PCA)  (Rao, 1964, Billard et al. 2023) and canonical analysis (CCA)   (Box and Tiao, 1977, Samadi et al., 2017) to achieve dimension reduction and improve predictions.
The asymptotic properties of the reduced-rank (RR) and the ordinary least squares (OLS) estimators of the VAR model are studied by  Anderson (2002).  
The   RRVAR model has been extended and combined with other approaches in economic and financial modeling, including,  common features and  RRVAR models (Franchi and Paruolo, 2011; Centoni and Cubadda, 2015; Cubadda et al., 2019), structural analysis through RRVAR models   (Bernardini and Cubadda, 2015; Carriero et al., 2016), Bayesian RRVAR models (Carriero et al., 2011; Cubadda and Guardabascio, 2019),  partial least squares approach (Cubadda and Hecq, 2011), the vector error correction model (VECM) under cointegration (L\"{u}tkepohl, 2005, Part II; Hecq et al., 2006), multivariate autoregressive index models (MAI) (Cubadda et al., 2017; Cubadda and Guardabascio, 2019),  heteroskedastic VAR models  (Hetland et al., 2021), time-varying parameter RRVAR models (Brune et al.,  2022). A detailed review of RRVAR models is given in Cubadda and Hecq (2022a).  These models are distinctly different from envelope models because there is no link between the mean function and covariance matrix (see \eqref{eq:2.4.1} in Section \ref{Envelope VAR model}).

While the reduced-rank VAR model achieves an effective dimensionality reduction, in many applications there are additional structures that can be exploited to achieve even higher dimensionality reduction with lower statistical error. The response envelope  
proposed by Cook et al. (2010) is another parsimonious approach to achieve dimension reduction and improve the estimation efficiency and prediction accuracy of standard multivariate regression models by parsimoniously decreasing the number of parameters. This method is useful in eliminating immaterial information present in the responses and predictors. The envelope method is effective even when the coefficient matrix is full rank, whereas the reduced-rank approach offers no reduction in this case. This is because the envelope uses the smallest reducing subspace of the covariance matrix that contains the mean function. There are several extensions of the basic envelope methodology to other contexts (Su and Cook, 2011; Cook et al., 2013; Cook and Zhang, 2015a, 2015b; Cook et al., 2015;    Su et al., 2016; Li and Zhang, 2017; Ding and Cook, 2018; Forzani and Su, 2021, Lee and Su, 2020).

Wang and Ding (2018) extended the envelope regression model proposed by Cook et al. (2010) to the VAR model called the envelope VAR (EVAR) model. The EVAR model provides better performance and is more efficient by removing immaterial information from estimation. Rekabdarkolaee et al. (2020) proposed a spatial envelope model for spatially correlated data in multivariate spatial regression. Samadi and DeAlwis (2023b) introduced the envelope matrix autoregressive (MAR) model (for the MAR model, refer to  Samadi, 2014).
 Cook et al. (2015) proposed a new parsimonious multivariate regression model by combining Anderson's (1999) RRR model with Cook et al.'s (2010) envelope model,  called the reduced-rank 
 envelope model.   
We incorporate the idea of envelopes (Cook et al., 2010; Wang and Ding, 2018) into Velu et al.'s (1986) RRVAR model and propose a novel efficient parsimonious VAR model for high-dimensional time series data. The proposed reduced-rank envelope VAR model combines the advantages and strengths of both the RRVAR and EVAR models which leads to more accurate estimation and higher efficiency.

We use the following notations and definitions throughout this paper. All real $k \times s$ matrices are denoted as $\R^{k \times s}$. The collection of $u$-dimensional subspaces in a $q$-dimensional vector space is called the Grassmannian manifold, indicated by $\mathcal{G}_{q, u}$. If $\bA \in \R^{k \times s}$, then $\Span(\bA) \subseteq \R^k$ is defined to be the subspace spanned by the columns of $\bA$. If $\sqrt{T}(\widehat{\boldsymbol\theta} - \boldsymbol\theta) \xrightarrow[]{\mathcal{D}} \mathcal{N}(\mathbf{0}, \bV)$, then the asymptotic covariance matrix of $\sqrt{T}\widehat{\boldsymbol\theta}$ is denoted as $\avar(\sqrt{T}\widehat{\boldsymbol\theta}) = \bV$. The projection matrix onto $\Span(\bX)$ in the $\bV$ inner product is denoted by $\bP_{\scriptsize\bX(\bV)} = \bX(\bX'\bV\bX)^{-1}\bX'\bV$, and the projection matrix onto $\Span(\bX)$ in the  identity inner product is denoted as $\bP_{\scriptsize\bX} = \bX(\bX'\bX)^{-1}\bX'$. Let $\bQ_{\scriptsize\bX(\bV)} = \bI - \bP_{\scriptsize\bX(\bV)}$, and $\bQ_{\scriptsize\bX} = \bI - \bP_{\scriptsize\bX}$. The operator $\vect: \R^{k \times s} \rightarrow \R^{ks}$  stacks the columns of a matrix into a column vector, and the operator $\vech: \R^{k \times k} \rightarrow \R^{k(k+1)/2}$ stacks the lower triangular part, including the diagonal, into a vector. Moreover, for any $q \times q$ symmetric matrix $\mathbf{U}$,  the expansion matrix  $\mathbf{E}_q \! \in \! \R^{q^{2} \times q(q+1)/2}$ is defined  such that   $\vect(\mathbf{U}) = \mathbf{E}_q\vech(\mathbf{U})$, and the contraction matrix $\mathbf{C}_q \in \R^{q(q+1)/2 \times q^{2}}$ is defined    such that  $\vech(\mathbf{U}) = \mathbf{C}_q\vect(\mathbf{U})$ (Henderson and Searle, 1979).   The Moore-Penrose generalized inverse of   $\mathbf{E}_q$  is defined as $\mathbf{E}^\dagger_q= (\mathbf{E}_q'\mathbf{E}_q)^{-1} \mathbf{E}_q'$. 
The Kronecker product of two matrices $\bX$ and $\bY$ is denoted by $\bX \otimes \bY$, and the symbol `$\sim$' means identically distributed. 

 The rest of the paper is structured as follows. Section \ref{Reduced-rank envelope vector autoregression} provides a comprehensive review of the reduced-rank VAR and the envelope VAR models, along with the introduction of our proposed approach, the reduced-rank envelope VAR (REVAR) model. The derivation of the maximum likelihood estimators (MLEs) for the parameters of the REVAR model is presented in  Section \ref{Likelihood estimation for REVAR model}.  
 In  Section \ref{Asymptotic Properties},  we establish the asymptotic properties of the proposed REVAR estimators with and without normality assumption and compare them with those obtained from the standard VAR, RRVAR, and EVAR models. The algorithms for selecting the lag order, rank, and envelope dimensions are outlined in Section \ref{Selections of lag order, rank, and envelope VAR dimension}. To assess the performance of our proposed REVAR model under different error assumptions, Section \ref{Simulation} presents the results of extensive simulation studies, comparing it with the reduced-rank VAR, envelope VAR, and standard VAR models. Real economic datasets are analyzed in Section \ref{Real data analysis}, and our conclusions are summarized in Section \ref{Discussion}. The Supplementary Materials contain the proofs of lemmas and propositions, as well as additional simulations.

\vspace{-.2in}

\section{Reduced-rank Envelope Vector Autoregression}\label{Reduced-rank envelope vector autoregression}
\vspace{-.2in}
\subsection{Reduced-rank VAR Model}\label{Reduced-rank VAR model}

 Suppose the coefficient matrix in model \eqref{eq:2.2} is rank deficient, i.e.,  $\rank(\bbeta)= d < q$.   As a result, it can be written as a product of two lower dimensional matrices, i.e., $ \bbeta = \mathbf{A}\mathbf{B}$.  Then, the reduced-rank VAR (RRVAR) model proposed by Velu et al. (1986) is given by 
\begin{equation}\label{eq:2.3}
 \mathbf{y}_t =\balpha +  \mathbf{A}\mathbf{B} \mathbf{x}_{t} + \boldsymbol\varepsilon_t,
\end{equation}
where $\mathbf{A} \in \mathbb{R}^{q \times d}$, $\mathbf{B} = [\mathbf{B}_{1}, \mathbf{B}_{2}, ..., \mathbf{B}_{p}] \in \mathbb{R}^{d \times qp}$, and $\rank(\mathbf{A}) =\rank(\mathbf{B})=d$.

The conditional log-likelihood function of model \eqref{eq:2.2} under the assumption of normality of $\boldsymbol\varepsilon_t$ can be written as (the initial values $(\by_{-p+1}, ..., \by_0)$ are assumed to be given)
\begin{equation}\label{eq:2.4}
  L_T(\balpha, \bbeta, \bSigma) \propto \frac{-T}{2}\{\log|\bSigma| + \frac{1}{T}\sum_{t=1}^{T}(\mathbf{y}_{t} - \balpha -\bbeta\mathbf{x}_{t})^{'}\bSigma^{-1}(\mathbf{y}_{t} - \balpha-\bbeta\mathbf{x}_{t})\}.
\end{equation}
Then, the maximization of $L_T(\balpha, \bbeta, \bSigma)$ is performed under the condition that rank$(\bbeta)=d$, or correspondingly under the RR autoregression parameterization $\bbeta = \mathbf{A}\mathbf{B}$ in model \eqref{eq:2.3}. The symbol $\propto$ is used to exclude unnecessary additive constants from the likelihood function. 
Before we present the MLEs of the RRVAR model \eqref{eq:2.3}, we introduce some notations.

Let $\boldsymbol{\Sigma}=\mbox{cov}(\boldsymbol\varepsilon_t)$, and let $\bGamma_k=\mbox{cov}(\by_{t-k}, \by_{t})$, $k \geq 0$, denotes the autocovariance matrix function of $\by_t$. Also, let $\bGamma_* = \mbox{cov}(\bx_t, \by_{t}) \in \R^{qp \times q}$ and $\bGamma_{(p)} = \mbox{cov}(\bx_t) \in \R^{qp \times qp}$ given as
\begin{equation*}
\bGamma_* = %\mbox{cov}(\bx_t,\by_t) =
\begin{pmatrix}[.7]
\bGamma_1\\
\bGamma_2\\
\vdots\\
\bGamma_{p}\\
\end{pmatrix},~~~
\bGamma_{(p)} = %\mbox{cov}(\bx_t,\bx_t) =
\begin{pmatrix}[.7]
\bGamma_0 & \bGamma_1' & \dots & \bGamma_{p-1}'\\
\bGamma_1 & \bGamma_0 & \dots & \bGamma_{p-2}'\\
\vdots & \vdots & \ddots & \vdots\\
\bGamma_{p-1} & \bGamma_{p-2} & \dots & \bGamma_0\\
\end{pmatrix}.
\end{equation*}

Given a series of observations $\by_{-p+1}, ..., \by_0, \by_1, ..., \by_T$, their sample covariance matrices are denoted by $\widehat{\bGamma}_*$ and  $\widehat{\bGamma}_{(.)}$ with the divisor $T$. Without loss of generality, suppose the sample lagged vector  $\mathbf{x}_{t}$ is centered, therefore $\bar{\mathbf{x}}={\bf 0}$.  
 For ease of notations, we shall drop the subsubscript ``$t$" in $\by_t$ and $\bx_t$ when they are subscripted. Let $\widehat{\bGamma}_k = \frac{1}{T}\sum_{t=1}^{T}(\mathbf{y}_{t-k} - \bar{\mathbf{y}})(\mathbf{y}_{t} - \bar{\mathbf{y}})'$ ($k = 0 ,1, \hdots$), and $\widehat{\bGamma}_{(p)} = \frac{1}{T}\sum_{t=1}^{T}\mathbf{x}_{t}\mathbf{x}_{t}'$ be the sample autocovariance matrices of $\by_t$ and $\bx_t$ at lag $k$, respectively; and $\widehat{\bGamma}_* = \frac{1}{T}\sum_{t=1}^{T}\mathbf{x}_{t}(\mathbf{y}_{t}-\bar{\mathbf{y}})'$ be the sample cross-covariance matrix between $\mathbf{x}_t$ and $\mathbf{y}_{t}$.
Moreover, let $\widehat{\bGamma}_{\mathbf{y}|\mathbf{x}}= \widehat{\bGamma}_0-\widehat{\bGamma}_*'\widehat{\bGamma}_{(p)}^{-1}\widehat{\bGamma}_*$ be the sample covariance matrix of residuals, 
and $\widehat{\bGamma}_{\mathbf{y}\circ\mathbf{x}} = \widehat{\bGamma}_*'\widehat{\bGamma}_{(p)}^{-1}\widehat{\bGamma}_*$ be the sample cross-covariance matrix of the fitted vectors, resulting from a vector autoregression of $\mathbf{y}_t$ on $\mathbf{x}_{t}$. The sample canonical correlation matrix between $\mathbf{y}_t$ and $\mathbf{x}_{t}$ is defined as $\mathbf{C}_{\mathbf{y},\mathbf{x}}  = \!\widehat{\bGamma}_0^{-1/2}\widehat{\bGamma}_*'\widehat{\bGamma}_{(p)}^{-1/2}$, where $\mathbf{C}_{\mathbf{x},\mathbf{y}}  =  \mathbf{C}_{\mathbf{y},\mathbf{x}}'$. Truncated matrices are denoted by superscripts, e.g.,  
$\mathbf{C}_{\mathbf{y},\mathbf{x}}^{(d)}$ and $\widehat{\bGamma}_*^{(d)}$ are formed by the $d$ eigenvectors associated with   the $d$ largest singular values of $\mathbf{C}_{\mathbf{y},\mathbf{x}}$ and $\widehat{\bGamma}_*$, respectively.

 Note that the decomposition $\bbeta = \bA\bB$ is not unique, as it can be expressed as  $\bbeta = \bA\mathbf{C}^{-1}\mathbf{C}\bB $    for any nonsingular matrix $\mathbf{C} \in \mathbb{R}^{q\times q}$.   The MLEs of the RRVAR model parameters, 
 obtained by Reinsel and Velu (1998), involve normalization constraints for identifiability, 
such as $\mathbf{B}\boldsymbol\bGamma_{(p)}\mathbf{B}^{'} = \boldsymbol\Lambda^2$, $\mathbf{A}^{'}\boldsymbol\rho\mathbf{A} = \mathbf{I}_d$,  where $\boldsymbol\Lambda^2 = \diag(\lambda_1^2, ..., \lambda_d^2 )$ and $\lambda_i^2$s are the  eigenvalues of  $\boldsymbol\rho^{1/2}\bGamma_{*}'\bGamma_{(p)}^{-1}\bGamma_{*}\boldsymbol\rho^{1/2}$, for any positive definite matrix $\boldsymbol\rho$. However, the parameters of interest $\bbeta$ and $\bSigma$ are identifiable, with  $ \Span(\bA)= \Span{(\bbeta)}$, $\Span(\bB')=\Span{(\bbeta')}$.  We propose a new framework for ML  
 estimation of $\bbeta$ and $\bSigma$ that does not require     constraints on   $\mathbf{A} $ or  $\mathbf{B} $, and  works for any  decomposition  $\bbeta = \bA\bB$ with $\rank(\mathbf{A}) =\rank(\mathbf{B})=d$.

The MLEs of the parameters of the  RRVAR model  
\eqref{eq:2.3}, which  maximize   \eqref{eq:2.4}, 
 are associated with the canonical correlations (Anderson, 2002), and  given by   $\widehat{\balpha}_{\RRVAR} = \bar{\mathbf{y}}$ and 
\begin{equation}\label{eq:2.4.2}
  \widehat{\bbeta}_{\RRVAR} = \widehat{\bGamma}_0^{1/2}\mathbf{C}_{\mathbf{y},\mathbf{x}}^{(d)} \widehat{\bGamma}_{(p)}^{-1/2}, ~~~~~
  \boldsymbol{\widehat{\Sigma}}_{\RRVAR} = \widehat{\bGamma}_0- \widehat{\bbeta}_{\RRVAR}\widehat{\bGamma}_* = \widehat{\bGamma}_0^{1/2}(\bI_q - \mathbf{C}_{\mathbf{y},\mathbf{x}}^{(d)}\mathbf{C}_{\mathbf{x},\mathbf{y}}^{(d)})\widehat{\bGamma}_0^{1/2}.
\end{equation}

There are different estimators for the RR parameters  ${\mathbf{A}}$   and ${\mathbf{B}}$ in the literature based on different constraints on
${\mathbf{A}}$   and ${\mathbf{B}}$.  All of them can   be reproduced by decomposing the rank-$d$ estimator  $\widehat{\boldsymbol\beta}_{\RRVAR}$ given in \eqref{eq:2.4.2}.  
The  OLSVAR estimators of $\bbeta$ and $\bSigma$ can be obtained by replacing the truncated sample canonical correlation matrices $\bC_{(.)}^{(d)}$ with their untruncated versions, i.e., $ \widehat{\boldsymbol{\beta}}_{\OLSVAR} = \widehat{\bGamma}_0^{1/2}\mathbf{C}_{\mathbf{y,x}}\widehat{\bGamma}_{(p)}^{-1/2}$ and $ \widehat{\boldsymbol{\Sigma}}_{\OLSVAR} = \widehat{\bGamma}_0^{1/2}(\mathbf{I}_{q} - \mathbf{C}_{\mathbf{y,x}} \mathbf{C}_{\mathbf{x,y}})\widehat{\bGamma}_0^{1/2}$.

\vspace{-.1in}

\subsection{Envelope VAR Model}\label{Envelope VAR model}

 The envelope model is a parsimonious model introduced by Cook et al. (2010) for multivariate regression that achieves efficiency gains in estimation and better prediction performance. An envelope model for time series data has been developed by Wang and Ding (2018) in the context of VAR  models. Before proceeding further, we introduce the following definitions which will be used in the following sections (see also Cook et al., 2010, 2015).
\begin{Definition}\label{def 2.1}
A subspace $\mathcal{R} \subseteq \mathbb{R}^{q}$ is defined to be a reducing subspace of $\mathbf{M} \in \mathbb{R}^{q \times q}$ ($\mathcal{R}$ reduces $\mathbf{M}$), if and only if $\mathbf{M} = \mathbf{P}_{\mathcal{R}}\mathbf{M}\mathbf{P}_{\mathcal{R}} + \mathbf{Q}_{\mathcal{R}}\mathbf{M}\mathbf{Q}_{\mathcal{R}}$.
\end{Definition}
The envelope methodology is based on the notion of reducing subspace, which is essential to the development of theory and practice in functional analysis (Conway, 1990).  

\begin{Definition}\label{def 2.2}
Let  $\mathbf{M} \in \mathbb{R}^{q \times q}$ and $\mathcal{S} \subseteq \emph{\Span}(\mathbf{M})$. Then the $\mathbf{M}$-envelope of $\mathcal{S}$ is defined as the intersection of all reducing subspaces of $\mathbf{M}$ that contains $\mathcal{S}$ and is denoted by $\mathcal{E}_{\mathbf{M}}(\mathcal{S})$.
\end{Definition}

Since the intersection of any two reducing subspaces of $\mathbf{M}$ is again a reducing subspace of $\mathbf{M}$,  Definition \ref{def 2.2} ensures the existence and uniqueness of envelopes.  For ease in notation,
we use $\mathcal{E}_{\mathbf{M}}(\mathcal{D})$ instead of $\mathcal{E}_{\scriptsize \mathbf{M}}\{\Span(\bD)\}$.  
It can be characterized that the $\mathbf{M}$-envelope is the span of some subset of the eigenvectors of $\mathbf{M}$ (Cook et al., 2010).  

 Let $\mathcal{S} \subseteq \R^{q}$ such that (i) $ \mathbf{Q}_\mathcal{S}\mathbf{y}_t\mid \mathbf{x}_{t} \sim \mathbf{Q}_\mathcal{S}\mathbf{y}_t$, and (ii) %$\mathbf{Q}_\mathcal{S}\mathbf{y}_t \! \! \indep \mathbf{P}_\mathcal{S}\mathbf{y}_t\mid \mathbf{x}_{t}$, 
 $ \cov\left( \mathbf{Q}_\mathcal{S}\mathbf{y}_t, \mathbf{P}_\mathcal{S}\mathbf{y}_t\mid \mathbf{x}_{t} \right) =0$,  
 where $\mathbf{P}_\mathcal{S}$ is the projection matrix onto $\mathcal{S}$, and $\mathbf{Q}_\mathcal{S}$ is the orthogonal projection. 
 $\mathbf{P}_\mathcal{S}\mathbf{y}_t$ and $\mathbf{Q}_\mathcal{S}\mathbf{y}_t$ are called material and immaterial parts of $\by_t$ respectively. 
 Conditions (i) and (ii) together imply that any dependence of $\mathbf{y}_t$ on its lagged values  $\mathbf{x}_t$ must be concentrated in the material part of the autoregression, i.e., $\mathbf{P}_\mathcal{S}\mathbf{y}_t$,  while    $\mathbf{Q}_\mathcal{S}\mathbf{y}_t$  is invariant to changes in $\mathbf{x}_t$ and becomes white noise and immaterial to the estimation of $\bbeta$. 
 Therefore, $\mathbf{P}_\mathcal{S}\mathbf{y}_t$ carry all material information in $\by_t$, and a change in $\mathbf{x}_t$  can affect the distribution of $\mathbf{y}_t$ only via $\mathbf{P}_\mathcal{S}\mathbf{y}_t$.
 If $\bvarepsilon_t$ is normally distributed, then condition (ii) implies that  $\mathbf{P}_\mathcal{S}\mathbf{y}_t$ is independent of  $\mathbf{Q}_\mathcal{S}\mathbf{y}_t$ given  $\mathbf{x}_{t}$.
 Under the envelope model, it can be shown that (i) and (ii) hold if and only if   (a) $\Span(\bbeta) = \mathcal{B}\subseteq \mathcal{S}$, and (b) $\mathcal{S}$ is a reducing subspace of $\bSigma$, i.e., $\bSigma =   \mathbf{P}_\mathcal{S}\bSigma \mathbf{P}_\mathcal{S}+\mathbf{Q}_\mathcal{S}\bSigma \mathbf{Q}_\mathcal{S}$ (Cook et al., 2010). That is, the   $\mathcal{S}$ is a reducing subspace of $\bSigma$ that contains $\mathcal{B}$, and the intersection of all such subspaces is called $\bSigma$-envelope of $\mathcal{B}$, and is denoted as $\mathcal{E}_{\scriptsize \bSigma}(\mathcal{B})$. Let $u = \dim(\mathcal{E}_{\scriptsize \mathbf{\bSigma}} (\mathcal{B}))$, then under the RRVAR model  \eqref{eq:2.3}, $\mathcal{E}_{\scriptsize {\bSigma}}(\mathcal{B}) = \mathcal{E}_{\scriptsize \mathbf{\bSigma}}(\mathcal{A})$, where $\mathcal{A}= \Span(\bA)$ and $u \geq d$. This is because $\dim\left(\mathcal{E}_{\scriptsize \mathbf{\bSigma}}(\mathcal{B})\right) \geq \dim(\mathcal{B}) = \rank(\bbeta) = d$.

Let $\bPhi \in \R^{q \times u}$ be an orthogonal basis of $\mathcal{E}_{\scriptsize \mathbf{\bSigma}}(\mathcal{B})$ with an orthogonal complement basis $\bPhi_0 \! \!\in \!\R^{q \times (q-u)}$ such that $(\bPhi, \bPhi_0)$ is an orthogonal matrix. Since $\Span(\bbeta) \subseteq \mathcal{E}_{\scriptsize \mathbf{\bSigma}}(\mathcal{B}) = \Span(\bPhi)$, there exists a $\boldsymbol\xi  \!   \in  \! \mathbb{R}^{u \times qp}$ such that $\bbeta = \bPhi\boldsymbol\xi$. %Moreover, 
Since $\mathcal{E}_{\scriptsize \mathbf{\bSigma}}(\mathcal{B})$ is spanned by a subset of     $\bSigma$'s eigenvectors, there exist symmetric positive definite matrices $\boldsymbol\Omega \in \mathbb{R}^{u \times u}$ and  $\boldsymbol\Omega_0 \in \mathbb{R}^{(q-u) \times (q-u)}$ such that $\bSigma = \bPhi\boldsymbol\Omega\bPhi^{'}+\bPhi_{0}\boldsymbol\Omega_{0} \bPhi_{0}^{'}$. That is, $\boldsymbol\xi \in \mathbb{R}^{u \times qp}$ contains the coordinates of $\bbeta$ in terms of the basis matrix $\bPhi$; and $\bOmega$ and $\bOmega_0$ carry the coordinates of  $\bSigma$ relative to $\bPhi$ and $\bPhi_0$, respectively. 
Thus, the $u$-dimensional envelope model is summarized as
\begin{equation}\label{eq:2.4.1}
 \bbeta =  \bPhi\boldsymbol\xi,~~ \bSigma = \bPhi\boldsymbol\Omega\bPhi^{'}+\bPhi_{0}\boldsymbol\Omega_{0} \bPhi_{0}^{'}.
\end{equation}
This provides a link between $\bbeta$ and $\bSigma$. That is, the white noise variation is decomposed into the variation related to the material part of $\by_t$, i.e., $\bPhi\boldsymbol\Omega\bPhi^{'}$ = Var$(\bP_{\scriptsize\bPhi}\by_t)$, and the variation related to the immaterial part of $\by_t$, i.e., $\bPhi_0\boldsymbol\Omega_0\bPhi_0^{'}$ = Var$(\bP_{\scriptsize\bPhi_0}\by_t)$. These two are respectively material and immaterial to the estimation of $\bbeta$.

Lam et al. (2011) and Lam and Yao (2012) developed 
  an approach that uses information from lagged autocovariance matrices via eigendecomposition to estimate the factor loading space. This approach has recently been extended to VAR models by Cubadda and Hecq (2022a, 2022b), and Wang et al. (2022b).  Although it follows that the orthogonal linear combinations of the factors convey all the relevant information for the estimation (like $\boldsymbol\Phi^{'} \mathbf{y}_t$ in the   EVAR model), and projection onto its orthogonal complement (like $\boldsymbol\Phi_0^{'} \mathbf{y}_t$ in the EVAR model)  gives white noise,   it is distinctly different from the   EVAR model as there is no required connection between    $\boldsymbol\beta$ and the error covariance matrix  $\boldsymbol\Sigma$.  The main difference between the two approaches is that the envelope methodology exploits connections between $\boldsymbol\beta$ and  $\boldsymbol\Sigma$ using a minimal reducing subspace, and the information of the basis matrix $\boldsymbol\Phi$ comes from both the conditional mean and the structure of covariance matrix.
 
%%%%%%%%%%%%%%%%%%%%%%%%%%%%%%%%%%%%%%%%%%%%%%%%%%%%%%%%%%%%%%%%%%%%%%%%%%%%%%%
\vspace{-.1in}

\subsection{Reduced-rank Envelope VAR Model}\label{Reduced-rank envelope VAR model}

 We propose a new parsimonious model, called the reduced-rank envelope vector autoregressive model (REVAR) model, by incorporating the idea of envelope VAR models into the reduced-rank VAR model.  This model improves the accuracy and efficiency of  VAR estimation.
 Given that $\Span(\bPhi) = \mathcal{E}_{\scriptsize \mathbf{\bSigma}}(\mathcal{B})$ 
 with $\dim(\mathcal{E}_{\scriptsize \mathbf{\bSigma}}(\mathcal{B})) = u \leq q$, the RRVAR model \eqref{eq:2.3} can be reparameterized into an envelope structure as
 \begin{equation}\label{eq:2.5}
 \bbeta = \mathbf{A}\mathbf{B} =  \bPhi\boldsymbol\xi =  \bPhi\bnu\mathbf{B},~~ \bSigma = \bPhi\boldsymbol\Omega\bPhi^{'}+\bPhi_{0}\boldsymbol\Omega_{0} \bPhi_{0}^{'},
\end{equation}
where the parameterization $\bbeta = \bPhi\boldsymbol\xi$ with $\boldsymbol\xi \in \mathbb{R}^{u \times qp}$ forms the EVAR model \eqref{eq:2.4.1} proposed by Wang and Ding (2018), and the columns of $\bnu \in \mathbb{R}^{u \times d}$, $u \geq d$, are the coordinates of $\mathbf{A}$ with respect to the basis $\bPhi$. Then the REVAR model is summarized as follows
\begin{equation}\label{eq:2.6}
\mathbf{y}_t = \balpha + \bPhi\bnu\mathbf{B} \mathbf{x}_{t}+\boldsymbol\varepsilon_t,~~ \bSigma = \bPhi\boldsymbol\Omega\bPhi^{'}+\bPhi_{0}\boldsymbol\Omega_{0} \bPhi_{0}^{'},
\end{equation}
 where $\boldsymbol\Omega$, $\boldsymbol\Omega_0$ are the same as those in \eqref{eq:2.4.1}.  There are  no further constraints on $\mathbf{A}, \mathbf{B}, \mbox{or}~\bnu$ in this model except that all  three matrices have rank $d$. Notice that the constrained parameters $\bA, \bB, \bPhi, \bPhi_0, \bxi, \bnu, \bOmega, \mbox{and}~ \bOmega_0$ in RRVAR, EVAR, and REVAR models are not unique, however, $\bbeta$ and $\bSigma$ in \eqref{eq:2.5} are unique.
In order to compare the properties of the REVAR model with the RRVAR model, we assume $\bPhi$ is known.
\begin{lem}\label{lem2}
Suppose $\bPhi$ is known, then the  maximum likelihood estimators (MLEs) of  the parameters of the REVAR model \eqref{eq:2.6}, which maximize \eqref{eq:2.4} are as follows,
\begin{equation*}
\begin{split}
  \widehat{\balpha}_{\scriptsize\bPhi} &= \bar{\mathbf{y}},~~~~~~ \widehat{\bbeta}_{\scriptsize\bPhi} =  \bPhi\boldsymbol{\widehat{\bnu}}_{\scriptsize\bPhi}\mathbf{\widehat{B}}_{\scriptsize\bPhi} =   \bPhi\widehat{\bGamma}_{\scriptsize\bPhi'\mathbf{y}}^{1/2}\mathbf{C}_{\scriptsize\bPhi'\mathbf{y},\mathbf{x}}^{(d)} \widehat{\bGamma}_{(p)}^{-1/2}\\
  \boldsymbol{\widehat{\Sigma}}_{\scriptsize\bPhi} &= \bPhi\widehat{\bGamma}_{\scriptsize\bPhi'\mathbf{y}}^{1/2}\left(\bI_u - \mathbf{C}_{\scriptsize\bPhi'\mathbf{y},\mathbf{x}}^{(d)}\mathbf{C}_{\scriptsize\mathbf{x},\bPhi'\mathbf{y}}^{(d)}\right)\widehat{\bGamma}_{\scriptsize\bPhi'\mathbf{y}}^{1/2}\bPhi' + \mathbf{Q}_{\scriptsize\bPhi}\widehat{\bGamma}_0\mathbf{Q}_{\scriptsize\bPhi}.
\end{split}
\end{equation*}
\end{lem}
 Note that the MLEs obtained in Lemma \ref{lem2} are based on $\bPhi$, however, the final estimators $\widehat{\bbeta}_{\scriptsize\bPhi}$ and $\boldsymbol{\widehat{\Sigma}}_{\scriptsize\bPhi}$ depend on $\bPhi$ only through $\Span(\bPhi)=\mathcal{E}_{\scriptsize \mathbf{\bSigma}}(\mathcal{B})$. That is, for any orthogonal matrix $\mathbf{O}\in \R^{u \times u}$ we have $\widehat{\bbeta}_{\scriptsize\bPhi} = \widehat{\bbeta}_{\scriptsize\bPhi\mathbf{O}}$ and $\boldsymbol{\widehat{\Sigma}}_{\scriptsize\bPhi} = \boldsymbol{\widehat{\Sigma}}_{\scriptsize\bPhi\mathbf{O}}$.

As a result, when the envelope is known, we should concentrate on the reduced response time series $\bPhi'\mathbf{y}_{t}$ and obtain the rank-$d$ RRVAR estimator $\boldsymbol{\widehat{\bnu}}_{\scriptsize\bPhi}\mathbf{\widehat{B}}_{\scriptsize\bPhi}$ of the vector autoregression of $\bPhi'\mathbf{y}_{t}$ on $\mathbf{x}_{t}$. Since $\widehat{\boldsymbol{\Sigma}}_{\scriptsize\bPhi} = \mathbf{P}_{\scriptsize\bPhi}\widehat{\boldsymbol{\Sigma}}_{\scriptsize\bPhi} \mathbf{P}_{\scriptsize\bPhi} + \mathbf{Q}_{\scriptsize\bPhi}\widehat{\boldsymbol{\Sigma}}_{\scriptsize\bPhi} \mathbf{Q}_{\scriptsize\bPhi}$, therefore according to Definition \ref{def 2.1}, the estimator $\widehat{\boldsymbol{\Sigma}}_{\scriptsize\bPhi}$ of  
$\boldsymbol{\Sigma}_{\scriptsize \bPhi}$ is now reduced by $\Span(\bPhi)$. Thus, $\Span(\boldsymbol\bPhi)$ is a reducing subspace of $\widehat{\boldsymbol{\Sigma}}_{\scriptsize\bPhi}$ that also includes $\Span(\widehat{\boldsymbol{\beta}}_{\scriptsize\bPhi})$, that is, the envelope subspace is maintained by the structure of the given ML estimators of the coefficient matrix. We obtain the ML estimator $\widehat{\boldsymbol{\Phi}}$ and show that the REVAR estimators for the coefficient matrix $\bbeta$ and covariance matrix $\bSigma$ are identical to the estimators in Lemma \ref{lem2} by substituting $\bPhi$ with $\widehat{\bPhi}$.

In comparing the models, notice that if the envelope dimension $u$ is equal to the response dimension, i.e.,   $ d<u = q$,     
then there is no immaterial information to be removed by the envelope VAR model and the REVAR model collapses to the RRVAR model \eqref{eq:2.3}, i.e., $\bPhi = \mathbf{I}_{q}$ and $\bA = \bnu$. If $d = u$, then the REVAR model is equivalent to the EVAR model. When $d=u=q$, the REVAR model collapses to the standard VAR model.  There is an extreme situation when $p > q = 1$, then both the RRVAR and EVAR models collapse to the standard univariate time series model. In this situation, there would not be any reduction. If $q > p = 1$, then both the RRVAR and REVAR models gain efficiency.  
Extreme cases in the REVAR model are numerically explored in Supplement S8.4 to support these assertions.
 
\vspace{-.15in}

\section{Estimation}\label{Likelihood estimation for REVAR model}

\vspace{-.2in}
\subsection{Parameterization for each Model}\label{Parameter comparison}

We define the following estimable functions for each model. Let $\mathbf{h}$ represent the parameter vector of the standard VAR model \eqref{eq:2.1}, and let $\boldsymbol\psi$, $\boldsymbol\delta$, and $\boldsymbol\theta$ denote the parameter vectors of the RRVAR, EVAR, and REVAR models, respectively. Since the estimator of the common parameter $\balpha$ is $\widehat{\balpha} = \bar{\mathbf{y}}$ for all methods, and $\bar{\mathbf{y}}$ is asymptotically independent of the estimators of $\bbeta$ and $\bSigma$; therefore $\balpha$ is omitted from all models. Hence, we have
\begin{equation}\label{eq:2.6.1}
\mathbf{h} =
\begin{pmatrix}[.75]
\vect{(\bbeta)}\\
\vech(\bSigma)
\end{pmatrix},~~~
\boldsymbol\psi =
\begin{pmatrix}[.75]
\vect{(\mathbf{A})}\\
\vect{(\mathbf{B})}\\
\vech(\bSigma)
\end{pmatrix},~~~
\boldsymbol \delta =
\begin{pmatrix}[.75]
\vect(\bPhi)\\
\vect(\boldsymbol\xi)\\
\vech(\boldsymbol\Omega)\\
\vech(\boldsymbol\Omega_0)
\end{pmatrix},~~~
\boldsymbol\theta =
\begin{pmatrix}[.7]
\vect(\bPhi)\\
\vect(\bnu)\\
\vect{(\mathbf{B})}\\
\vech(\boldsymbol\Omega)\\
\vech(\boldsymbol\Omega_0)
\end{pmatrix},
\end{equation}
where $\mathbf{h} = (\mathbf{h}_{1}', \mathbf{h}_{2}')'$, $\boldsymbol\psi = (\boldsymbol\psi_1', \boldsymbol\psi_2', \boldsymbol\psi_3')'$, $\boldsymbol \delta = (\boldsymbol\delta_1', \hdots, \boldsymbol\delta_4')'$, and $\boldsymbol\theta = (\boldsymbol\theta_1', ..., \boldsymbol\theta_5')'$, respectively. 
 Note that $\mathbf{h} = \mathbf{h}(\boldsymbol\psi)$ for the RRVAR model, $\mathbf{h} = \mathbf{h}(\boldsymbol\delta)$ for the EVAR model, and $\mathbf{h} = \mathbf{h}(\boldsymbol\theta)$ for the REVAR model.
To compare the models, we use $\mathcal{T}(\cdot)$ to represent the total number of parameters (NOP) in $\mathbf{h}, \boldsymbol\psi, \boldsymbol \delta$, and $\boldsymbol\theta$. Then, the parameter count for each model is
\begin{enumerate}[itemsep=-7pt, label=(\roman*)]  %[(i)]
  \item Standard VAR($p$) model, $\mathcal{T}_{\OLSVAR} = \mathcal{T}(\mathbf{h})= q^{2}p + q(q+1)/2$;
  \item Reduced-rank VAR($p$) model, $\mathcal{T}_{\RRVAR} = \mathcal{T}(\boldsymbol\psi) = d\{q(p+1)-d\}+q(q+1)/2$;
  \item Envelope VAR($p$) model, $\mathcal{T}_{\EVAR} = \mathcal{T}(\boldsymbol\delta) = uqp + q(q+1)/2$;
  \item Reduced-rank envelope VAR($p$) model, $\mathcal{T}_{\REVAR} = \mathcal{T}(\boldsymbol\theta)  = d\{qp+u-d\}+ q(q+1)/2 $.
\end{enumerate}

The reduced NOP from the standard VAR model to the RRVAR model is $\mathcal{T}(\mathbf{h}) - \mathcal{T}(\boldsymbol\psi) = (qp-d)(q-d) \geq 0$, and that is even further reduced from the RRVAR model to the REVAR model by the number of  $\mathcal{T}(\boldsymbol\psi) - \mathcal{T}(\boldsymbol\theta) = d(q-u) \geq 0$. Moreover, compared to the standard VAR model, the reduced NOP by the   EVAR model is $\mathcal{T}(\mathbf{h}) - \mathcal{T}(\boldsymbol\delta) = qp(q-u) \geq 0$, and it is further reduced by $\mathcal{T}(\boldsymbol\delta) - \mathcal{T}(\boldsymbol\theta) = (qp-d)(u-d) \geq 0$ from the   EVAR model to the REVAR model.  
It is important to note that envelope models achieve efficiency gains not only from parsimony but mainly from the structure of the covariance matrix.
To evaluate the impact of covariance matrix structure and variations in the immaterial and material matrices on the performance of envelope models, we conducted simulations by adjusting the ratios of immaterial variation to material variation, as presented in Supplement S8.5.

\vspace{-.1in}

\subsection{Maximum Likelihood Estimation for the REVAR Model}\label{Estimators for the REVAR model parameters}

In this section, we derive the ML estimators of the   REVAR model \eqref{eq:2.6} for a given lag order $p$, rank $d$, and envelope dimension $u$. How to select $p$, $d$, and $u$ are discussed in Section \ref{Selections of lag order, rank, and envelope VAR dimension}.  
The MLEs for the REVAR model can be obtained by replacing  $\mathbf{h}$ with $\mathbf{h}(\boldsymbol\theta)$ in the log-likelihood function \eqref{eq:2.4} as $L_T \left(\balpha, \bbeta(\boldsymbol\theta), \bSigma(\boldsymbol\theta)\right) \equiv L_T(\balpha, \mathbf{B}, \bnu, \boldsymbol\Omega, \boldsymbol\Omega_0, \bPhi | p, d, u)$ and maximizing it with respect to all parameters other than $\bPhi$. Since $\Span(\bPhi)$ lies in a Grassmann manifold (Cook et al., 2010), therefore we cannot analytically find a unique optimal value of $\bPhi$. As it is discussed in Proposition \ref{prop2}, we can obtain $\widehat{\bPhi}$ from minimization over a Grassmannian.

To obtain the MLE of $\bPhi$, let $\mathbf{D} \in \R^{q \times u}$ be a semiorthogonal matrix and define the standardized version of $\mathbf{D}'\mathbf{y}_t \in \mathbb{R}^{u}$ as $\mathbf{z}_{\mathbf{D}} = (\mathbf{D}'\widehat{\bGamma}_0\mathbf{D})^{-1/2}\mathbf{D}'\mathbf{y}_t$ with sample covariance matrix $\mathbf{I}_{u}$. Suppose $\widehat{\omega}_i(\mathbf{D})$, $i = 1, \hdots, u$, is the $i$th eigenvalue of $\widehat{\bGamma}_{\mathbf{z}_{\mathbf{D}}|\mathbf{x}}^{-1}$ which is also the $i$th eigenvalue of $(\mathbf{D}'\widehat{\bGamma}_{\mathbf{y|x}}\mathbf{D})^{-1/2}(\mathbf{D}'\widehat{\bGamma}_0\mathbf{D})(\mathbf{D}'\widehat{\bGamma}_{\mathbf{y|x}}\mathbf{D})^{-1/2}$.

\begin{proposition}\label{prop2}
Let $L_T(\balpha, \mathbf{B}, \bnu, \boldsymbol\Omega, \boldsymbol\Omega_0, \bPhi | p, d, u)$ denotes the log-likelihood function of the  REVAR model \eqref{eq:2.6}.  Then,  it achieves its maximum at $\widehat{\bPhi} = \arg \min_{\mathbf{D}} \mathbf{F}_{T}(\mathbf{D}|p,d,u)$, where
\begin{align}
   \mathbf{F}_{T}(\mathbf{D}|p,d,u) &= \log|\mathbf{D}'\widehat{\bGamma}_0\mathbf{D}|+\log|\mathbf{D}'\widehat{\bGamma}_0^{-1}\mathbf{D}|+\log|\mathbf{I}_u - \widehat{\bGamma}^{(d)}_{\mathbf{z}_{\mathbf{D}}\circ\mathbf{x}}| \label{eq:2.7}\\
   &= \log|\mathbf{D}'\widehat{\bGamma}_{\mathbf{y|x}}\mathbf{D}|+\log|\mathbf{D}'\widehat{\bGamma}_0^{-1}\mathbf{D}|+\sum_{i=d+1}^{u}\log\{\widehat{\omega}_i(\mathbf{D})\},\label{eq:2.8}
\end{align}
where the minimization is over the Grassmannian of dimension $u \in \R^{q}$, i.e., $\mathcal{G}_{q,u}$.
\end{proposition}

The expansion of the objective function in \eqref{eq:2.7} offers a method of interpreting the log-likelihood functions for envelope-based models. For instance, the objective  function of the EVAR model \eqref{eq:2.4.1} (Wang and Ding, 2018) can be similarly expanded as follows
\begin{equation}\label{eq:2.9}
\log|\mathbf{D}'\widehat{\bGamma}_0\mathbf{D}|+\log|\mathbf{D}'\widehat{\bGamma}_0^{-1}\mathbf{D}|+\log|\mathbf{I}_u - \widehat{\bGamma}_{\mathbf{z}_{\mathbf{D}}\circ\mathbf{x}}|,
\end{equation}
which is similar to the expression \eqref{eq:2.7} except that the last term in \eqref{eq:2.9} is based on the ordinary least squares method, whereas the last term in \eqref{eq:2.7} is based on the reduced-rank VAR model. In other words, the two expressions in \eqref{eq:2.7} and \eqref{eq:2.9} are the same when $d = u$.
\begin{proposition}\label{prop3} The sample objective function $\mathbf{F}_{T}(\mathbf{D}|p,d,u)$ in \eqref{eq:2.8} converges in probability to  its  population counterpart   $\mathbf{F} (\mathbf{D}|p,u) = \log|\mathbf{D}'\bSigma\mathbf{D}|+\log|\mathbf{D}'\bGamma_0^{-1}\mathbf{D}|$ uniformly in $\mathbf{D}$, as $T \rightarrow\infty$. The estimator $\widehat{\bPhi} = \arg \min_{\mathbf{D}} \mathbf{F}_{T}(\mathbf{D}|p,d,u)$ is consistent, and $\widehat{\mathcal{E}}_{\scriptsize\bSigma}(\mathcal{B}) = \textup{\Span}\{ \arg \min_{\mathbf{D}} \mathbf{F}_{T}(\mathbf{D}|p,u)\}$.
\end{proposition}

The following Proposition provides the MLEs of the REVAR model parameters in \eqref{eq:2.6}.

\begin{proposition}\label{prop4}
The ML estimators for the REVAR model \eqref{eq:2.6} that maximize the conditional log-likelihood function in \eqref{eq:2.4} are $\widehat{\balpha}_{\REVAR} = \bar{\mathbf{y}}$, $\widehat{\bPhi} = \arg \min_{\mathbf{D}} \mathbf{F}_{T}(\mathbf{D}|p,d,u)$, and
\begin{align*}
 \widehat{\boldsymbol\Omega}_0 &= \widehat{\bPhi}_0'\widehat{\bGamma}_0\widehat{\bPhi}_0, ~~~~~~
  \boldsymbol{\widehat{\Omega}}_{} = \widehat{\bGamma}_{\scriptsize \bPhi'\mathbf{y}}^{1/2}\left(\mathbf{I}_u - \mathbf{C}_{ \scriptsize \bPhi'\mathbf{y},\mathbf{x}}^{(d)}\mathbf{C}_{\mathbf{x}, {\scriptsize\bPhi'}\mathbf{y}}^{(d)}\right)\widehat{\bGamma}_{ \scriptsize \bPhi'\mathbf{y}}^{1/2}, \\
  \widehat{\bbeta}_{\REVAR} &= \widehat{\bPhi}\widehat{\bnu}\widehat{\mathbf{B}}_{\REVAR} = \widehat{\bPhi}\widehat{\bGamma}_{ \scriptsize \bPhi'\mathbf{y}}^{1/2}\mathbf{C}_{\scriptsize \bPhi'\mathbf{y},\mathbf{x}}^{(d)} \widehat{\bGamma}_{(p)}^{-1/2}, ~~~~
  \widehat{\bSigma}_{\REVAR} = \widehat{\bPhi}\widehat{\boldsymbol\Omega}\widehat{\bPhi}^{'}+\widehat{\bPhi}_{0}\widehat{\boldsymbol\Omega}_{0} \widehat{\bPhi}_{0}^{'}.
 \end{align*}
\end{proposition}
Note that  rank$(\widehat{\bbeta}_{\REVAR}) = d ~(\leq u)$, and $\Span(\widehat{\bbeta}_{\REVAR}) \subseteq \mathcal{E}_{\scriptsize \mathbf{\bSigma}}(\mathcal{B})$. In comparison to the 
RRVAR model,  $\widehat{\boldsymbol{\Sigma}}_{\REVAR} = \mathbf{P}_{ \scriptsize \widehat{\bPhi}}\widehat{\boldsymbol{\Sigma}}_{\REVAR} \mathbf{P}_{\scriptsize \widehat{\bPhi}} + \mathbf{Q}_{\scriptsize \widehat{\bPhi}}\widehat{\boldsymbol{\Sigma}}_{\REVAR} \mathbf{Q}_{\widehat{\scriptsize \bPhi}}$
now has an envelope structure.
If $u = q$, then $\bPhi = \bI_q$, $\bA = \bnu$, and the REVAR model is reduced to the RRVAR model.  

 The full Grassmann (FG) optimization method can be computationally slow and expensive, particularly in high-dimensional problems. Therefore, we used the  FG algorithm for the small VARs, but for higher dimensional cases, we adapt and employ the one-dimensional (1D) algorithm proposed by Cook and Zhang (2016).  
  The 1D algorithm is computationally more efficient and robust compared to the FG optimization. This is due to the fact that it decomposes the $u$-dimensional FG optimization into a series of $u$  one-dimensional optimization problems and does not require an initial guess (see    Cook and Zhang, 2018).

\vspace{-.2in}

\section{Asymptotic Properties}\label{Asymptotic Properties}
\vspace{-.15in}

In this section, we establish the asymptotic properties of the proposed ML estimators assuming both Gaussian and non-Gaussian white noise processes. 
We derive their asymptotic distributions 
under the reduced-rank envelope VAR model and compare their asymptotic efficiencies to those of the OLSVAR, RRVAR, and REVAR models.  
 Related asymptotic results for estimators of the parameters of the  
 RRVAR, EVAR, and OLSVAR models 
 can be found in Anderson (2002),  Wang and Ding (2018), and  L\"{u}tkepohl (2005),  respectively.    
The asymptotic comparison between $\widehat{\bbeta}_{\RRVAR}$ and $\widehat{\bbeta}_{\OLSVAR}$ can be found in Anderson (2002).   Therefore,  our  focus here is  on comparing   $\widehat{\bbeta}_{\REVAR}$ and $\widehat{\bbeta}_{\RRVAR}$. The asymptotic results of $\widehat{\bbeta}_{\REVAR}$ over $\widehat{\bbeta}_{\EVAR}$ are comparable to those of $\widehat{\bbeta}_{\RRVAR}$ over $\widehat{\bbeta}_{\OLSVAR}$. This is due to the rank reduction constraint present in the material response time series $\bPhi'\by_t$.

Let $\mathbf{J}_\mathbf{h}$ be the  Fisher information of 
$\mathbf{h} = \left(\mathbf{h}_{1}', \mathbf{h}_{2}'\right)'= (\vect'(\bbeta), \vech'(\bSigma))'$. 
Then, the asymptotic covariance matrix of the OLSVAR estimator of $\mathbf{h}$ is  given by (L\"{u}tkepohl, 2005)
\begin{equation}\label{eq:9.0}
\emph{\avar}(\sqrt{T} \widehat{\mathbf{h}}_{\OLSVAR})=\mathbf{J}^{-1}_\mathbf{h}=\begin{pmatrix}[.8]
\mathbf{J}^{-1}_{\scriptsize \bbeta} &\mathbf{0}\\
\mathbf{0} &\mathbf{J}^{-1}_{\scriptsize \bSigma}\\
\end{pmatrix}=
\begin{pmatrix}[.8]
\bGamma^{-1}_{(p)} \otimes \bSigma &\mathbf{0}\\
\mathbf{0} & 2\mathbf{E}^\dagger_q(\bSigma \otimes \bSigma){\mathbf{E}^\dagger_q}'\\
\end{pmatrix}.
\end{equation}
This is also the asymptotic covariance matrix of the unrestricted ML estimator of $\mathbf{h}$.
Let us define the gradient matrices of $\mathbf{h}(\boldsymbol\psi)$ and $\mathbf{h}(\boldsymbol\theta)$ as
\begin{equation}\label{peq:9.1}
   \mathbf{H} = \frac{\partial\mathbf{h}(\boldsymbol\psi)}{\partial\boldsymbol\psi} =
\begin{pmatrix}[.8]
\frac{\partial\mathbf{h}_1}{\partial\boldsymbol\psi_1'} &\hdots &\frac{\partial\mathbf{h}_1}{\partial\boldsymbol\psi_3'}\\
%\vdots & &\vdots\\
\frac{\partial\mathbf{h}_2}{\partial\boldsymbol\psi_1'} &\hdots &\frac{\partial\mathbf{h}_2}{\partial\boldsymbol\psi_3'}\\
\end{pmatrix},~~~ \mathbf{R} = \frac{\partial\mathbf{h}(\boldsymbol\theta)}{\partial\boldsymbol\theta} =
\begin{pmatrix}[.8]
\frac{\partial\mathbf{h}_1}{\partial\boldsymbol\theta_1'} &\hdots &\frac{\partial\mathbf{h}_1}{\partial\boldsymbol\theta_5'}\\
%\vdots & &\vdots\\
\frac{\partial\mathbf{h}_2}{\partial\boldsymbol\theta_1'} &\hdots &\frac{\partial\mathbf{h}_2}{\partial\boldsymbol\theta_5'}\\
\end{pmatrix}.
\end{equation}
The asymptotic covariance matrices of the RRVAR and REVAR estimators, i.e.,  $\widehat{\mathbf{h}}_{\RRVAR} = \mathbf{h}(\widehat{\boldsymbol\psi})$ and  $\widehat{\mathbf{h}}_{\REVAR} = \mathbf{h}(\widehat{\boldsymbol\theta})$,  are provided in Proposition \ref{prop5},  
which can be obtained by using the asymptotic theory of overparameterized structural models proposed by Shapiro (1986).

\begin{proposition}\label{prop5}
Suppose $\boldsymbol\varepsilon_t \sim \mathcal{N}(\mathbf{0}, \bSigma)$, then it can be shown that 
$\emph{\avar}(\sqrt{T} \widehat{\mathbf{h}}_{\OLSVAR}) = \mathbf{J}_{\mathbf{h}}^{-1}$, $\emph{\avar}(\sqrt{T}\widehat{\mathbf{h}}_{\RRVAR}) = \mathbf{H}(\mathbf{H}'\mathbf{J}_{\mathbf{h}}\mathbf{H})^{\dagger}\mathbf{H}'$, and $\emph{\avar}(\sqrt{T}\widehat{\mathbf{h}}_{\REVAR}) = \mathbf{R}(\mathbf{R}'\mathbf{J}_{\mathbf{h}}\mathbf{R})^{\dagger}\mathbf{R}'$. The differences between the asymptotic covariances are as follows
\begin{align*}
   \avar(\sqrt{T}\widehat{\mathbf{h}}_{\OLSVAR}) - \avar(\sqrt{T}\widehat{\mathbf{h}}_{\RRVAR}) &= \mathbf{J}_{\mathbf{h}}^{-1/2}\mathbf{Q}_{\mathbf{J}_{\mathbf{h}}^{1/2}\mathbf{H}}\mathbf{J}_{\mathbf{h}}^{-1/2}\geq 0, \\
  \avar(\sqrt{T}\widehat{\mathbf{h}}_{\RRVAR}) - \avar(\sqrt{T}\widehat{\mathbf{h}}_{\REVAR}) &= \mathbf{J}_{\mathbf{h}}^{-1/2}\left(\mathbf{P}_{\mathbf{J}_{\mathbf{h}}^{1/2}\mathbf{H}} - \mathbf{P}_{\mathbf{J}_{\mathbf{h}}^{1/2}\mathbf{R}}\right)\mathbf{J}_{\mathbf{h}}^{-1/2} \\
   &= \mathbf{J}_{\mathbf{h}}^{-1/2}\mathbf{P}_{\scriptsize \mathbf{J}_{\mathbf{h}}^{1/2}\mathbf{H}}\mathbf{Q}_{\mathbf{J}_{\mathbf{h}}^{1/2}\mathbf{R}}\mathbf{J}_{\mathbf{h}}^{-1/2} \geq 0,
\end{align*}
where $\dagger$ denotes the Moore-Penrose inverse. Moreover, we have  $\avar\{\sqrt{T}\vect(\widehat{\bbeta}_{\OLSVAR})\} \! \geq \avar\{\sqrt{T}\vect(\widehat{\bbeta}_{\RRVAR})\} \geq \avar\{\sqrt{T}\vect(\widehat{\bbeta}_{\REVAR})\}$. Similarly in compared to the envelope VAR estimator, $\avar\{\sqrt{T}\vect(\widehat{\bbeta}_{\OLSVAR})\} \geq \avar\{\sqrt{T}\vect(\widehat{\bbeta}_{\EVAR})\} \geq \avar\{\sqrt{T}\vect(\widehat{\bbeta}_{\REVAR})\}$.
\end{proposition}
 
\begin{proposition}\label{prop6}
Suppose $\boldsymbol\varepsilon_t \sim \mathcal{N}(\mathbf{0}, \bSigma)$, and the rank of the coefficient matrix $\bbeta$ is $d$. Then, $\sqrt{T}\emph{\vect}(\widehat{\bbeta}_{\OLSVAR} - \bbeta)$ and $\sqrt{T}\emph{\vect}(\widehat{\bbeta}_{\RRVAR} - \bbeta)$ asymptotically follow normal distributions with mean zero and the following covariance matrices
\begin{align}
  \emph{\avar}\{\sqrt{T}\emph{\vect}(\widehat{\bbeta}_{\OLSVAR})\} &= \bGamma_{(p)}^{-1} \otimes \bSigma, \nonumber \\
  \emph{\avar}\{\sqrt{T}\emph{\vect}(\widehat{\bbeta}_{\RRVAR})\} &= (\mathbf{I}_{q^2p} - \mathbf{Q}_{\scriptsize\mathbf{B}'(\bGamma_{(p)})} \otimes \mathbf{Q}_{\scriptsize\mathbf{A}(\bSigma^{-1})})\emph{\avar}\{\sqrt{T}\emph{\vect}(\widehat{\bbeta}_{\OLSVAR})\} \label{eq:2.10}\\
   &= \emph{\avar}\{\sqrt{T}\emph{\vect}(\widehat{\bbeta}_{\mathbf{A}}\mathbf{Q}'_{\scriptsize\mathbf{B}'(\bGamma_{(p)})})\}+\emph{\avar}\{\sqrt{T}\emph{\vect}(\widehat{\bbeta}_{\mathbf{B}})\},  \label{eq:2.11}
\end{align}
where $\emph{\avar}\{\sqrt{T}\emph{\vect}(\widehat{\bbeta}_{\mathbf{A}}\mathbf{Q}'_{\scriptsize\mathbf{B}'(\bGamma_{(p)})})\} = (\mathbf{Q}_{\scriptsize\mathbf{B}'(\bGamma_{(p)})}\bGamma_{(p)}^{-1})\otimes (\mathbf{P}_{\scriptsize\mathbf{A}(\bSigma^{-1})}\bSigma)$, and $\emph{\avar}\{\sqrt{T}\emph{\vect}(\widehat{\bbeta}_{\mathbf{B}})\} = (\mathbf{P}_{\scriptsize\mathbf{B}'(\bGamma_{(p)})}\bGamma_{(p)}^{-1})\otimes\bSigma$.
\end{proposition}

 The asymptotic variance in \eqref{eq:2.10} follows from Anderson (2002, Eq. 5.22).   Moreover, the asymptotic results in Proposition \ref{prop6} depend on  the decomposed components $\bA$ and $\bB$ only through their orthogonal projections $\mathbf{Q}_{\scriptsize\mathbf{A}(\bSigma^{-1})}$  and  $\mathbf{Q}_{\scriptsize\mathbf{B}(\bGamma_{(p)})}$, respectively. Therefore, these results are valid for any    decomposition  $\bbeta = \bA\bB$ that satisfies $\mathbf{A} \in \mathbb{R}^{q\times d}$ and $\mathbf{B} \in \mathbb{R}^{d\times qp}$.

\begin{proposition}\label{prop7}
Suppose $\boldsymbol\varepsilon_t \sim \mathcal{N}(\mathbf{0}, \bSigma)$, then under the REVAR model $\sqrt{T}\emph{\vect}(\widehat{\bbeta}_{\REVAR} - \bbeta)$ follows asymptotic  normal distribution with zero mean and covariance
\begin{align}
  \emph{\avar}\{\sqrt{T}\emph{\vect}(\widehat{\bbeta}_{\REVAR})\} &= \emph{\avar}\{\sqrt{T}\emph{\vect}(\widehat{\bbeta}_{\scriptsize\bPhi}) \} + \emph{\avar}\{\sqrt{T}\emph{\vect}(\mathbf{Q}_{\scriptsize\bPhi}\widehat{\bbeta}_{\scriptsize\bnu, \mathbf{B}}) \} \label{eq:2.12}\\[1ex]
  \begin{split}
  &= \emph{\avar}\{\sqrt{T}\emph{\vect}(\widehat{\bbeta}_{\scriptsize\bPhi, \bnu}\mathbf{Q}'_{\scriptsize\mathbf{B}'(\bGamma_{(p)})}) \} + \emph{\avar}\{\sqrt{T}\emph{\vect}(\widehat{\bbeta}_{\scriptsize\bPhi, \mathbf{B}}) \} \\
  &~~~~~~~~~~~~~~+ \emph{\avar}\{\sqrt{T}\emph{\vect}(\mathbf{Q}_{\scriptsize\bPhi}\widehat{\bbeta}_{\scriptsize\bnu, \mathbf{B}}) \},  \label{eq:2.13}
  \end{split}
\end{align}
where from \eqref{eq:2.11} we have   $\emph{\avar}\{\sqrt{T}\emph{\vect}(\widehat{\bbeta}_{\scriptsize\bPhi, \bnu}\mathbf{Q}'_{\scriptsize\mathbf{B}'(\bGamma_{(p)})}) \} = \emph{\avar}\{\sqrt{T}\emph{\vect}(\widehat{\bbeta}_{\scriptsize\mathbf{A}}\mathbf{Q}'_{\scriptsize\mathbf{B}'(\bGamma_{(p)})}) \}$.
\end{proposition}

The asymptotic advantages of the REVAR   over the RRVAR model can be obtained by using the results in Propositions \ref{prop6} and \ref{prop7}. By subtracting \eqref{eq:2.13} from  \eqref{eq:2.11}, we have  

\begin{align}\label{eq:2.14}
\begin{split}
&\emph{\avar}\{\sqrt{T}\emph{\vect}(\widehat{\bbeta}_{\RRVAR})\} - \emph{\avar}\{\sqrt{T}\emph{\vect}(\widehat{\bbeta}_{\REVAR})\} \\
 & ~~~~~~~~~~=  \avar\{\sqrt{T}\vect(\widehat{\bbeta}_{\scriptsize\bB})\} - \avar\{\sqrt{T}\vect(\widehat{\bbeta}_{\scriptsize\bPhi,\bB})\} - \avar\{\sqrt{T}\vect(\mathbf{Q}_{\scriptsize\bPhi}\widehat{\bbeta}_{\scriptsize \bnu, \mathbf{B}}) \} \geq 0,
  \end{split}
\end{align}
where for a given $\mathbf{B} \in \R^{d \times qp}$, the estimators are defined as $\widehat{\bbeta}_{\scriptsize\bB} = \widehat{\bA}_{\scriptsize\bB}\bB$,  $\ \widehat{\bbeta}_{\scriptsize\bPhi,\bB} = \bPhi\widehat{\bnu}_{\scriptsize\bPhi,\bB}~\bB = \widehat{\bA}_{\scriptsize\bPhi,\bB} \bB$, and $\widehat{\bbeta}_{\scriptsize\bnu,\bB} = \widehat{\bPhi}_{\scriptsize\bnu,\bB}~\bnu\bB = \widehat{\bA}_{\scriptsize\bnu,\bB}\bB$.

\vspace{-.2in}
%%%%%%%%%%%%%%%%%%%%%%%%%%%%%%%%%%%%%%%%%%%%%%%%%%
 
\subsection{Asymptotic Properties under Non-normality}    \vspace{-.1in}
Suppose $\widehat{\mathbf{h}}_{\OLSVAR} = \big(\vect'(\widehat{\bbeta}_{\OLSVAR}),  \vech'(\widehat{\bGamma}_{\mathbf{y|x}})\big)'$ is  the OLSVAR estimator of $\mathbf{h}$ under the unrestricted  VAR model, and $\widehat{\mathbf{h}}_{\REVAR} = \mathbf{h}(\widehat{\boldsymbol\psi})$ denote the REVAR estimator. Moreover, we assume that $\mathbf{h}_0$ and $\boldsymbol{\psi}_0$ are the true values of $\mathbf{h}$ and $\boldsymbol{\psi}$, respectively. The following objective function is obtained after maximizing $L_T(\balpha, \bbeta, \bSigma)$  in  \eqref{eq:2.4} with respect to $\balpha$
\begin{equation*}
  L_T(\bbeta, \bSigma) \propto \frac{-T}{2}\left\{\log|\bSigma| + \tr\left[\bSigma^{-1}\left(\widehat{\bGamma}_{\mathbf{y|x}} + (\widehat{\bbeta}_{\OLSVAR} - \bbeta)\widehat{\bGamma}_{(p)}(\widehat{\bbeta}_{\OLSVAR} - \bbeta)'\right)\right]\right\}.
\end{equation*}
We now consider $L_T(\bbeta, \bSigma)$ as a function of $\mathbf{h}$ and $\widehat{\mathbf{h}}_{\OLSVAR}$, and define the discrepancy function as   $\mathcal{F}(\mathbf{h}, \widehat{\mathbf{h}}_{\OLSVAR})  = 2/T\big\{ L_T(\widehat{\bbeta}_{\OLSVAR}, \widehat{\bGamma}_{\mathbf{y|x}}) - L_T(\bbeta, \bSigma) \big\}$, which satisfies the necessary conditions of Shapiro's (1986). It can be shown that the second derivative of $\mathcal{F}(\mathbf{h}, \widehat{\mathbf{h}}_{\OLSVAR})$, i.e., $\mathbf{J}_{\mathbf{h}} \! \! = 1/2\Big( \partial^2\mathcal{F}(\mathbf{h}, \widehat{\mathbf{h}}_{\OLSVAR})/\partial\mathbf{h}\partial\mathbf{h}' \Big)$  evaluated  at $\widehat{\mathbf{h}}_{\OLSVAR} \!\! = \mathbf{h}_0$ and $\mathbf{h} = \mathbf{h}_0$ is the Fisher information matrix for $\mathbf{h}$   when $\bvarepsilon_t$ is normally distributed. The following proposition gives the asymptotic distribution of $\widehat{\mathbf{h}}_{\REVAR}$ without the normality assumption on $\bvarepsilon_t$.

\begin{proposition}\label{prop8}
Assume that the error terms of the reduced-rank envelope VAR model \eqref{eq:2.6}  are independent and identically distributed (i.i.d)   and the fourth moments of $\bvarepsilon_t$ are finite. Then, $\sqrt{T}(\widehat{\mathbf{h}}_{\OLSVAR} - \mathbf{h}_0) \xrightarrow[]{\mathcal{D}}  \mathcal{N}(\mathbf{0}, \tilde{\bV})$ for some positive definite covariance matrix $\tilde{\bV}$, and $\sqrt{T}(\widehat{\mathbf{h}}_{\REVAR} - \mathbf{h}_0) \xrightarrow[]{\mathcal{D}} \mathcal{N}(\mathbf{0}, \tilde{\bZ})$, with $\tilde{\bZ} = \bR(\bR'\mathbf{J}_{\mathbf{h}}\bR)^{\dagger}\bR'\mathbf{J}_{\mathbf{h}} \tilde{\bV}\mathbf{J}_{\mathbf{h}}\bR(\bR' \mathbf{J}_{\mathbf{h}}\bR)^{\dagger}\bR'$,
where  $ \mathbf{J}_{\mathbf{h}}$ and $\bR$  are defined in \eqref{eq:9.0} and \eqref{peq:9.1}, respectively. Particularly, $\sqrt{T}(\vect(\widehat{\bbeta}_{\REVAR}) - \vect(\widehat{\bbeta})) \xrightarrow[]{\mathcal{D}}  \mathcal{N}(\mathbf{0}, \tilde{\bZ}_{11})$, where  $\tilde{\bZ}_{11}$   is the top-left block matrix of $\tilde{\bZ}$ of dimension $q^2p \times q^2p$.
\end{proposition}

The $\sqrt{T}$-consistency of $\widehat{\bbeta}_{\REVAR}$,  the reduced-rank envelope VAR estimator,  relies on the  $\sqrt{T}$-consistency of both $\widehat{\bbeta}_{\OLSVAR}$ and $\widehat{\bGamma}_{\mathbf{y|x}}$, despite the non-normality of the error terms, and the properties of the discrepancy function $\mathcal{F}(\mathbf{h}, \widehat{\mathbf{h}}_{\OLSVAR})$.  The asymptotic covariance matrix $\tilde{\bZ}_{11}$ can be estimated conveniently by plugging in the estimated covariance matrix $\tilde{\bV}$ into $\tilde{\bZ}$, but its accuracy depends on the distribution of $\bvarepsilon_t$ for any fixed sample size.

\vspace{-.1in}

\section{Selections of  $p$,   $d$, and     $u$ of VAR Models}\label{Selections of lag order, rank, and envelope VAR dimension}
\vspace{-.2in}

\subsection{Lag Order ($p$) Selection}\label{Lag order ($p$) selection}
 
The selection of the lag order ($p$) is an empirical problem and a critical element in the specification of VAR models.
Therefore, the first step in VAR analysis is to determine the lag order.  
To do this, several model selection criteria are employed. The common approach is to fit VAR($p$) models with lag orders $p = 0, \hdots, p_{\max}$ and select the value of $p$ that minimizes some model selection information criteria (IC). The general form of IC
for VAR models is given by $IC(p) = \ln\lvert\widehat{\bSigma}(p)\lvert + c_T \ \phi(p)$  (L\"{u}tkepohl, 2005; Tsay, 2014),
where $\widehat{\bSigma}(p) = T^{-1}\sum_{t=1}^{T}\widehat{\boldsymbol\varepsilon}_t\widehat{\boldsymbol\varepsilon}_t'$, which is the estimated residual covariance matrix for a model of order $p$, $c_T$ is a sequence that depends on the sample size $T$, $\phi(p)$ is a penalty function that penalizes large VAR($p$) models. The most common information criteria are the Akaike information criterion (AIC) and the Bayesian information criterion (BIC),  given as
\begin{align*}
  AIC(p) &= \ln\lvert\widehat{\bSigma}(p)\lvert + \frac{2}{T}\mathcal{T}_{\OLSVAR}, ~~~~
  BIC(p) = \ln\lvert\widehat{\bSigma}(p)\lvert + \frac{\ln(T)}{T}\mathcal{T}_{\OLSVAR},  
\end{align*}
where $\mathcal{T}_{\OLSVAR} = q^2p + q(q+1)/2$, is the number of parameters in the standard VAR model.

\vspace{-.1in}
\subsection{Rank ($d$) Selection}\label{Rank ($d$) selection}
\vspace{-.1in}
The implicit assumption of RR models is that the coefficient matrix is not of full rank. Optimal rank selection has an important role in dimension reduction. In the context of RRR, 
Bura and Cook (2003) proposed a rank selection test that follows a chi-squared distribution and only requires the finite second moments of the response variables. The rank of  
$\bbeta$ can be  determined using  
the test statistic  $\mathcal{M}_{d} = T\sum_{j=d+1}^{q}\lambda_j^2$,  
where $\lambda_1 \geq \hdots \geq \lambda_q$ are the eigenvalues of the   $qp \times q$ matrix  $ \widehat{\bbeta}_{\textup{std}}=\sqrt{(T-qp-1)/T}~\widehat{\bGamma}_{(p)}^{1/2}\widehat{\bbeta}_{\OLSVAR}'\widehat{\bGamma}_{\by|\bx}^{-1/2}$.
   Under the null hypothesis $H_0$: $d = d_0$,  the test statistic $ \mathcal{M}_{d_0}$ has an asymptotic chi-square distribution, that is, $\mathcal{M}_{d_0}\sim\mathcal{X}_{(qp-d_0)(q-d_0)}^2$ (Bura and Cook, 2003). 
   To find the rank $d$, one can compute 
   a sequence of test statistics $ \mathcal{M}_{d_0}$, for $d_0 = 0,1,\hdots, q-1$, and compare them to the percentiles of their corresponding null   
   distributions $\mathcal{X}_{(qp-d_0)(q-d_0)}^2$. The test procedure stops at the first nonsignificant test of $H_0$: $d = d_0$. Then, $d_0$ is an estimate of the rank of   $\bbeta$.

\vspace{-.1in}
\subsection{Envelope Dimension ($u$) Selection}\label{Envelope VAR dimension ($u$) selection}

 To select the envelope dimension $u$, there are two ways.
 One approach is to simultaneously determine  $d$ and  $u$  by seeking $(d, u)$ from $(0, 0)$ to $(q, q)$  ($d \leq u \leq q$), and selecting the    $(d, u)$ pair that has the lowest IC value. The  IC
  to determine the optimal $(d, u)$ pair  are 
\begin{align*}
  AIC(d, u) = 2 \mathcal{T}_{\REVAR} - 2\hat{L}_{d, u}, ~~~~~ BIC(d, u) = \log(T) \mathcal{T}_{\REVAR} - 2\hat{L}_{d, u},
\end{align*}
where $\mathcal{T}_{\REVAR}$ is the number of REVAR model parameters, and $\hat{L}_{d, u}$ is  the maximized log-likelihood function calculated at the MLEs  
in Proposition \ref{prop4}.
Alternatively,   we can first find $d$ using the method in Section \ref{Rank ($d$) selection},  then search for the value of  $u$ from $d$ to $q-1$ that minimizes AIC or BIC.
In this case, under the null hypothesis $H_0: u=u_0$,  the test statistic $\mathcal{M}_{d,u_0}= 2\big(\hat{L}_{d, q}- \hat{L}_{d, u_0}\big)$ has an asymptotic chi-square distribution with $(q-u_0)d$ degree of freedom, i.e,  $\mathcal{M}_{d,u_0} \sim\mathcal{X}_{(q-u_0)d}^2$.
The latter method is computationally more efficient.

%%%%%%%%%%%%%%%%%%%%%%%%%%%%%%%%%%%%%%%%%%%%%%%%%%%%%%%%%%%%%%%%%%%%%%%%%%%%%%%%%%%%%%%%%%%%%%%%%%%%%%%%%%%%%%%%%%%%%%%%%%%%%%%%%%%%%%%%%%%%%%%%%%%%%%%%%%%%%%%%%%%%%%%%%%%%%%%%%%%%%%%%
\vspace{-.25in}
\section{Simulation Studies}\label{Simulation}
\vspace{-.1in}
  
In this section, we compare the performance of our proposed REVAR model with the RRVAR, the EVAR model, and the OLSVAR model using simulation studies under different data generating processes (DGP).  
We simulate data under various parameter settings from model \eqref{eq:2.6} with $[\Omega]_{ij} = (-0.9)^{|i-j|}$ and $[\Omega_0]_{ij} = 5(-0.5)^{|i-j|}$ for $\bOmega$ and $\bOmega_0$, respectively. The semiorthogonal matrix $\bPhi$ and matrix $\bB$ are generated from the uniform distribution on $(0,1)$. The entries of $\bnu$ and $p$ presample observations ($\by_0, \hdots, \by_{-p+1}$) are generated from the standard normal distribution. Then, $\bPhi$ and $\bbeta = \bPhi\bnu\bB$ are standardized so that $\bPhi'\bPhi = \bI_u$ and $||\bbeta||_F = 1$, where $||.||_F$ is the Frobenius norm, and the coefficient matrix $\bbeta$ satisfies the stationary condition. Estimation errors are obtained by comparing the estimated coefficient matrix $\widehat{\bbeta}$ to the true coefficient matrix $\bbeta$ by using $||\bbeta - \widehat{\bbeta}||_F$. We conduct simulation studies under various error distributions with sample sizes of $T = 160,\, 270,\, 450, \, 740, \, 1200, \,\mbox{and}~2000$.  
Each scenario is replicated 100 times, and the minimum and maximum standard error ratios of coefficient estimates ($r_{\min}=\min ({SE_{\textup{M}}}/{SE_{\REVAR}})$  and $r_{\max}=\max ({SE_{\textup{M}}}/{SE_{\REVAR}})$) are calculated for the M=OLSVAR, EVAR, and RRVAR models compared to our proposed REVAR model.
  Matlab codes are available upon request.

\vspace{-.15in}
  \subsection{Simulation Studies with Normal Errors}\label{Simulation_normal}

\begin{figure}[H]%\label{fig1.a}
 %\vspace{-1.cm}
   \centering
 \begin{subfigure}[b]{0.5\textwidth}
  \includegraphics[width=\textwidth,height=4.6cm]{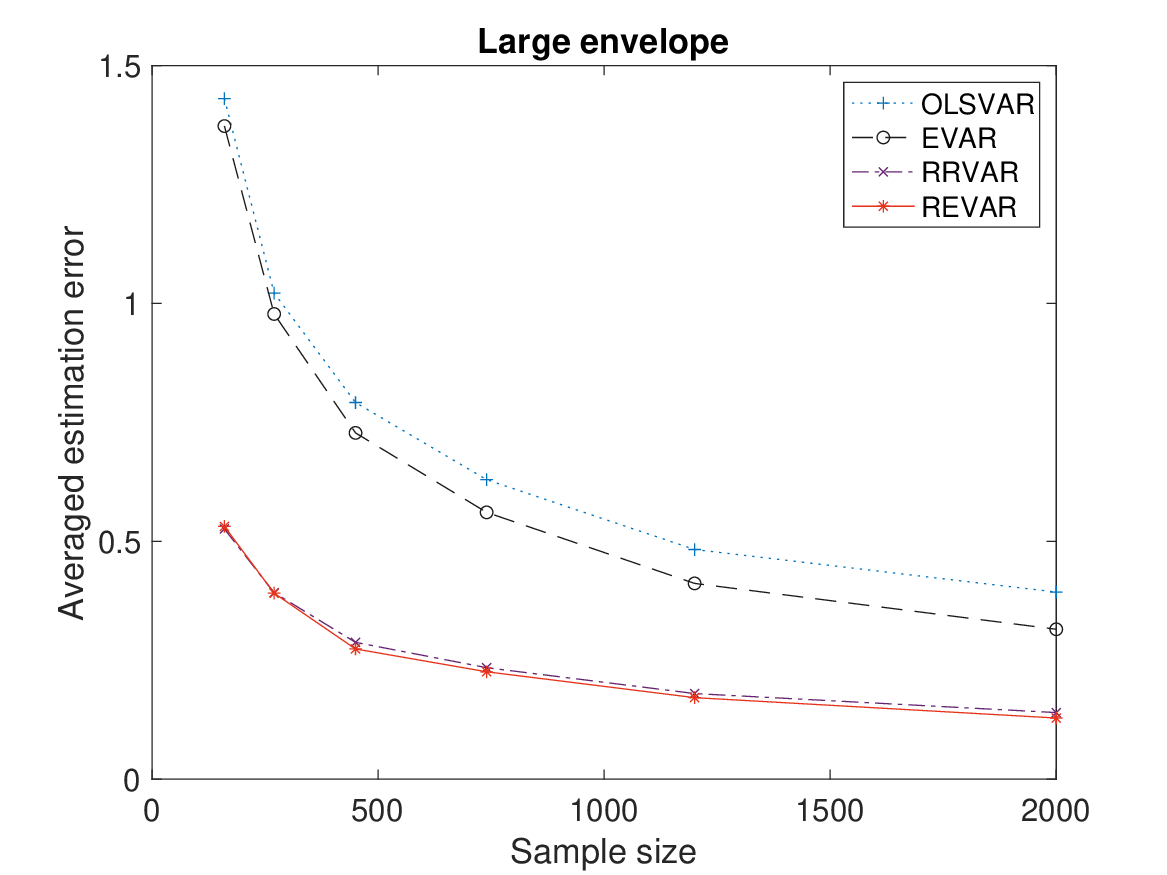}
                \caption{ $(d, u, p, q)=(2, 6, 1, 7)$}
        \end{subfigure}%
         \hspace{-.6cm}
        %add desired spacing between images, e. g. ~, \quad, \qquad etc.
   \begin{subfigure}[b]{0.5\textwidth}
         \includegraphics[width=\textwidth,height=4.6cm]{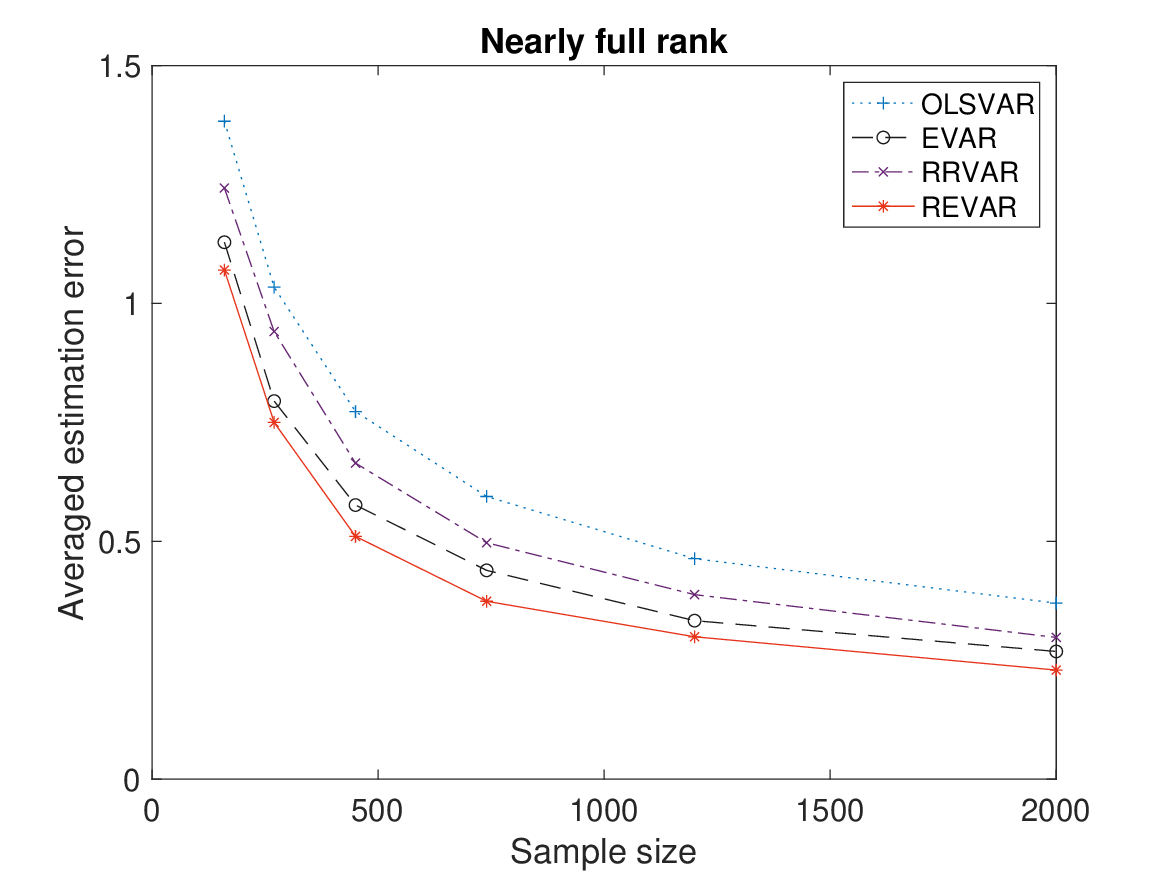}
                \caption{$(d, u, p, q)=(5, 6, 1, 7)$ }
        \end{subfigure}
    %\vspace{-.1cm}
        \centering
 \begin{subfigure}[b]{0.5\textwidth}
                \includegraphics[width=\textwidth,height=4.6cm]{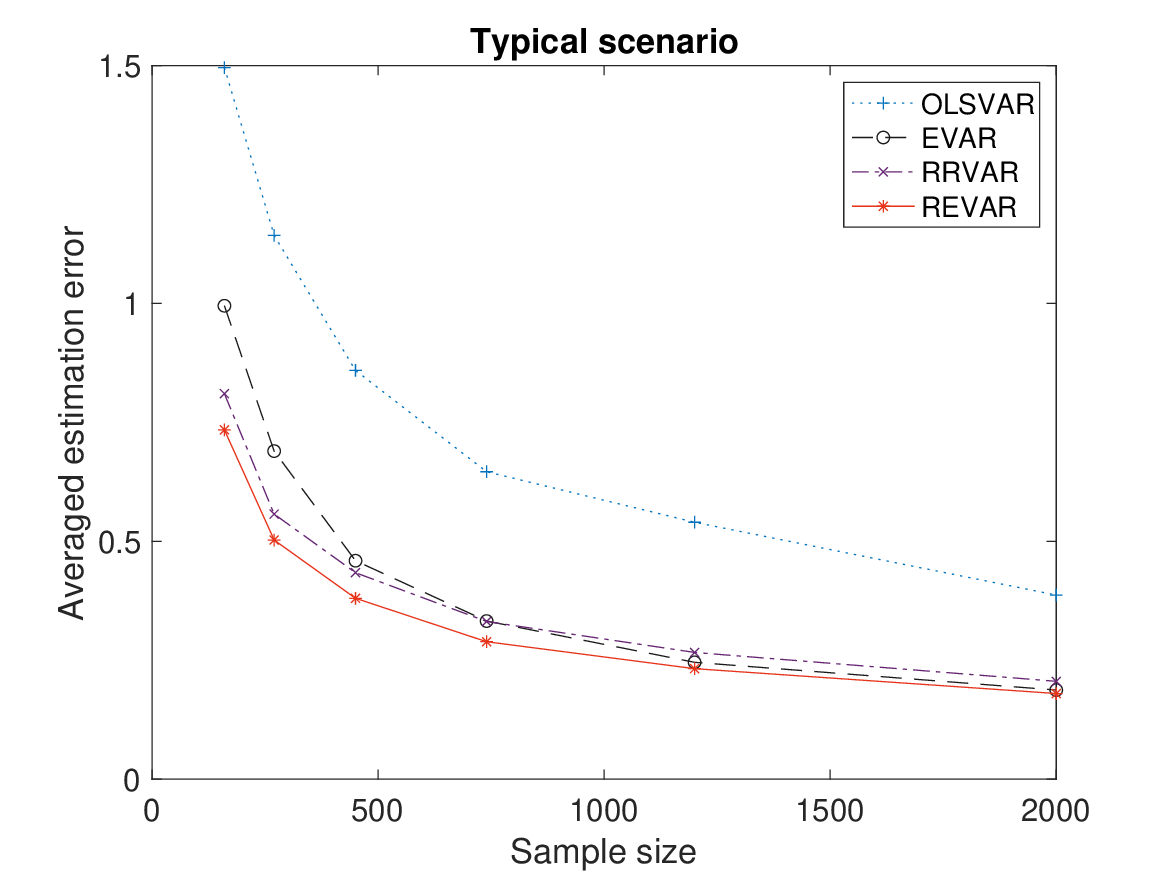}
                \caption{ $(d, u, p, q)=(3, 4, 1, 7)$}
        \end{subfigure}%
      \hspace{-.6cm}  %add desired spacing between images, e. g. ~, \quad, \qquad etc.
           \begin{subfigure}[b]{0.5\textwidth}
                \includegraphics[width=\textwidth,height=4.6cm]{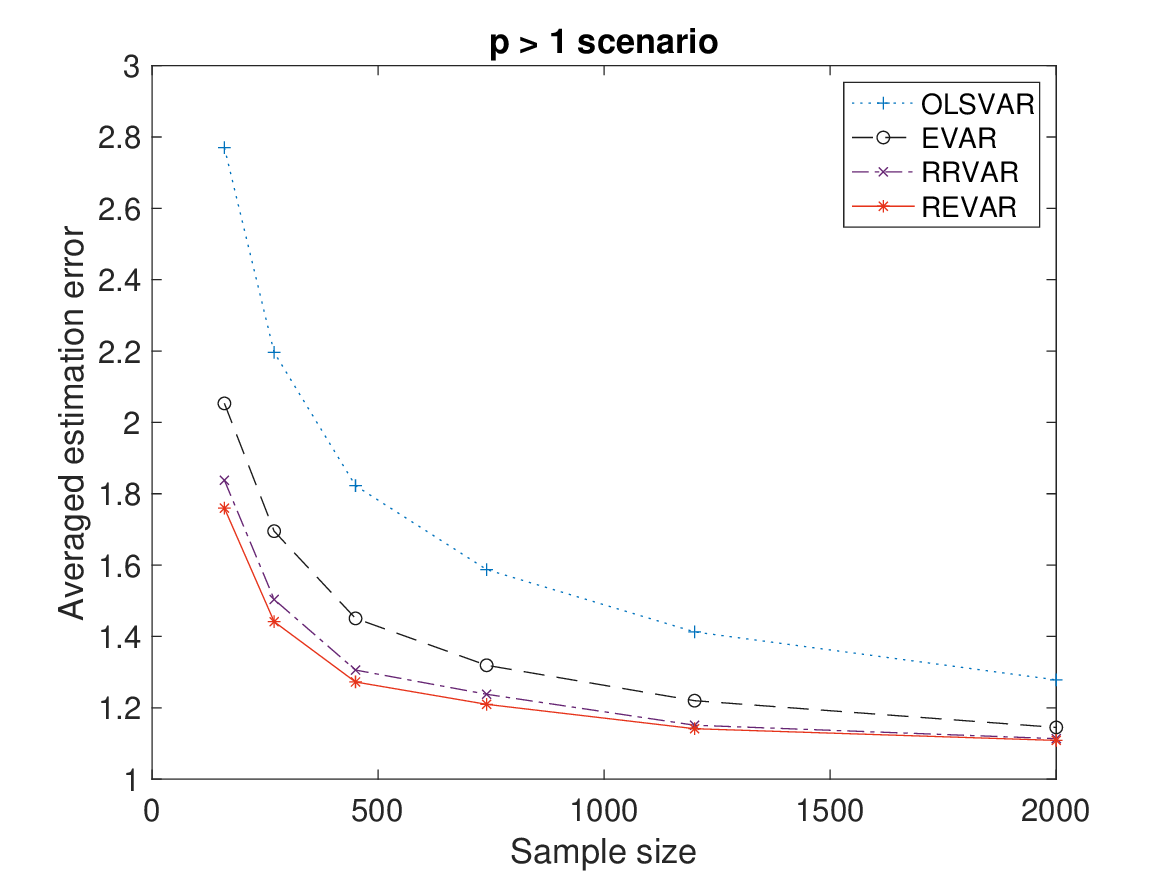}
               \caption{$(d, u, p, q)=(3, 4, 2, 7)$ }
        \end{subfigure}
     \vspace{.3cm}
    \caption{Average estimation errors of moderate-dimensional VAR models with different combinations of $(d, u, p, q)$   against sample size.  Refer to  Table \ref{tb:1} for details }

     % \caption{For each plots, from left to right, $(d, u, p, q)$ are $(2, 6, 1, 7), (5, 6, 1, 7), (3, 4, 1, 7)$, and $(3, 4, 2, 7)$. Sample sizes are  $160, 270, 450, 740, 1200, \,\mbox{and}\, 2000$.}
       \label{fig1}
   \end{figure}

  \vspace{-.05in}
In this subsection, data with normal errors are analyzed.
Table \ref{tb:1} shows the total number of parameters for different  VAR models in our simulation studies with various combinations of $(d, u, p, q)$.  The REVAR model has the fewest parameters, making it more parsimonious than other models. Consequently, it significantly outperforms them,  as shown in    Figure \ref{fig1}.

\begin{table}[H]
        \small
	\renewcommand{\arraystretch}{.52}
	\caption{Total number of parameters (NOP) in each VAR model correspond to Figure \ref{fig1} }
	\begin{center}
		\begin{tabular}{c c c c c c} \hline
		\centering
			      & &Large envelop  &Nearly full rank  &Typical Scenario &$p>1$\\
			        &  $(d, u, p, q)$ &  $(2, 6, 1, 7)$ &$(5, 6, 1, 7)$ & $(3, 4, 1, 7)$ &$(3, 4, 2, 7)$\\
			      \hline
				
			      &OLSVAR &77  &77    &77   &126\\
				  &RRVAR     &52 &73    &61    &82\\
                  &EVAR    &70 &70    &56    &84\\
				  &REVAR    &50  &68    &52        &73\\

				         \hline
		\end{tabular}
	\end{center}
	\label{tb:1}
\end{table}

\vspace{-.2cm}
 
  Figure \ref{fig1} displays the impact of envelope dimension and rank on the relative performance of each method.  %on each method's relative performance. 
  The simulation results in Figure \ref{fig1}  are obtained using the full Grassmannian (FG) algorithm.  
  In the scenario with a large envelope dimension (Figure  \ref{fig1}(a)),
  the RRVAR approach outperforms   
  the OLSVAR approach, while the  EVAR  model shows a relatively smaller improvement. The REVAR approach has a slight advantage over RRVAR in this scenario.  
  In the second case (Figure  \ref{fig1}(b)), where $\bbeta$ is almost full rank, the RRVAR method outperforms the OLSVAR  by a small margin. Both the EVAR and  REVAR  approaches demonstrate significant improvement over the RRVAR and  OLSVAR, where the  REVAR   is superior to all others.  For the third scenario (Figure  \ref{fig1}(c))  neither the  EVAR   nor RRVAR approaches are favored. Both the RRVAR and EVAR approaches exhibit similar behaviour and show substantial improvements compared to the OLSVAR approach.
  The final scenario (Figure  \ref{fig1}(d)) is similar to the third but with a higher number of lags $(p)$.

Our proposed  REVAR model combines the advantages and strengths of both reduced-rank and envelope VAR models,  demonstrating its superiority over alternative methods in simulation studies.  

\begin{figure}[!htbp]
 %\vspace{-1.cm}
   \centering
 \begin{subfigure}[b]{0.5\textwidth}
            \includegraphics[width=\textwidth,height=4.5cm]{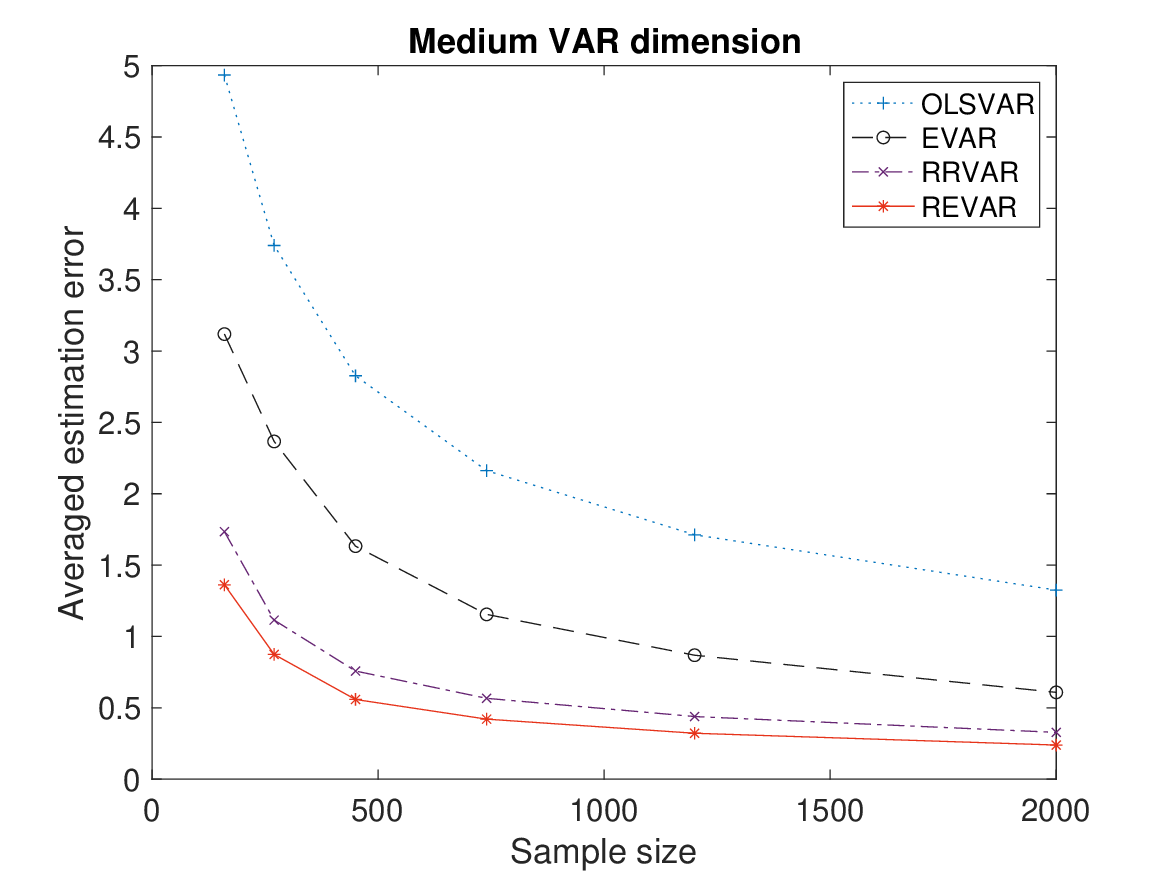}
               \caption{ $(d, u, p, q)=(5, 10, 1, 20)$}
        \end{subfigure}%
         \hspace{-.5cm}
           \begin{subfigure}[b]{0.5\textwidth}
            \includegraphics[width=\textwidth,height=4.5cm]{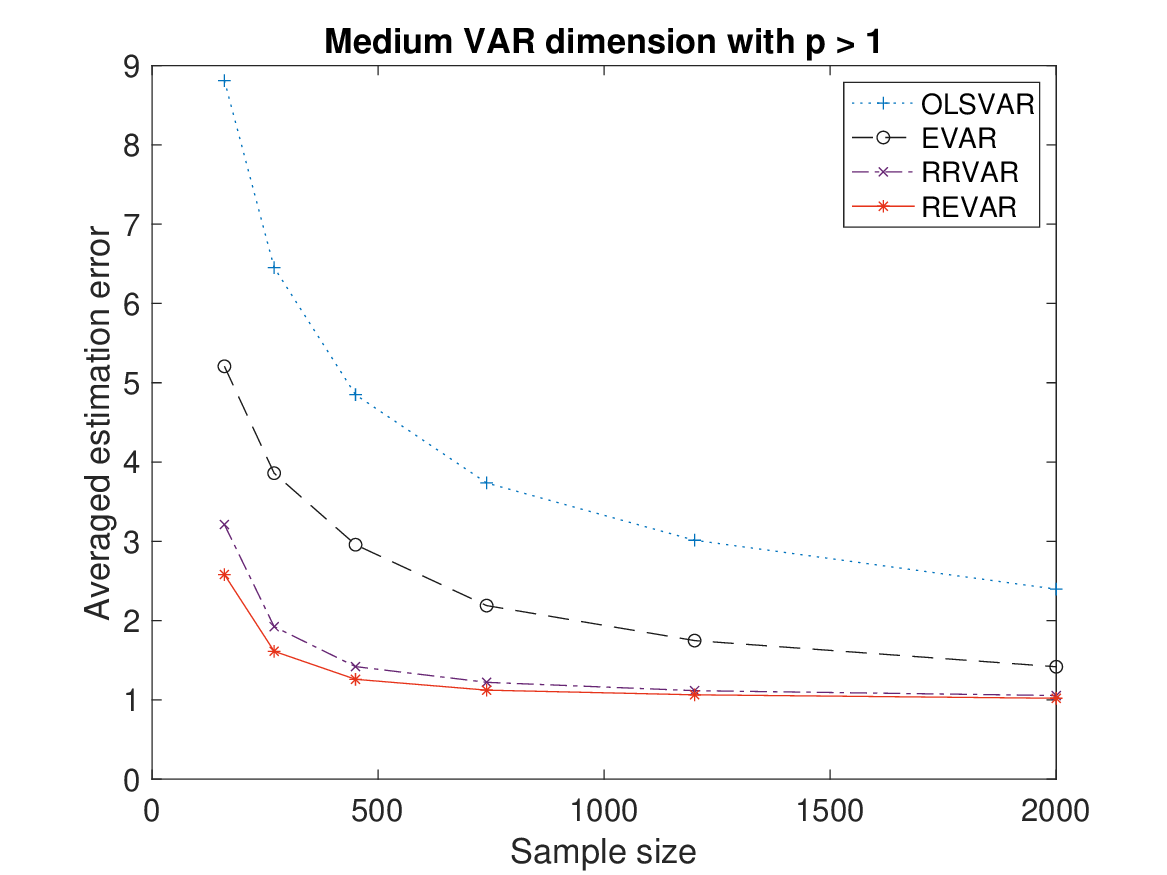}
              \caption{ $(d, u, p, q)=(5, 10, 2, 20)$}
        \end{subfigure}
    %\vspace{-.1cm}
        \centering
 \begin{subfigure}[b]{0.5\textwidth}
            \includegraphics[width=\textwidth,height=4.5cm]{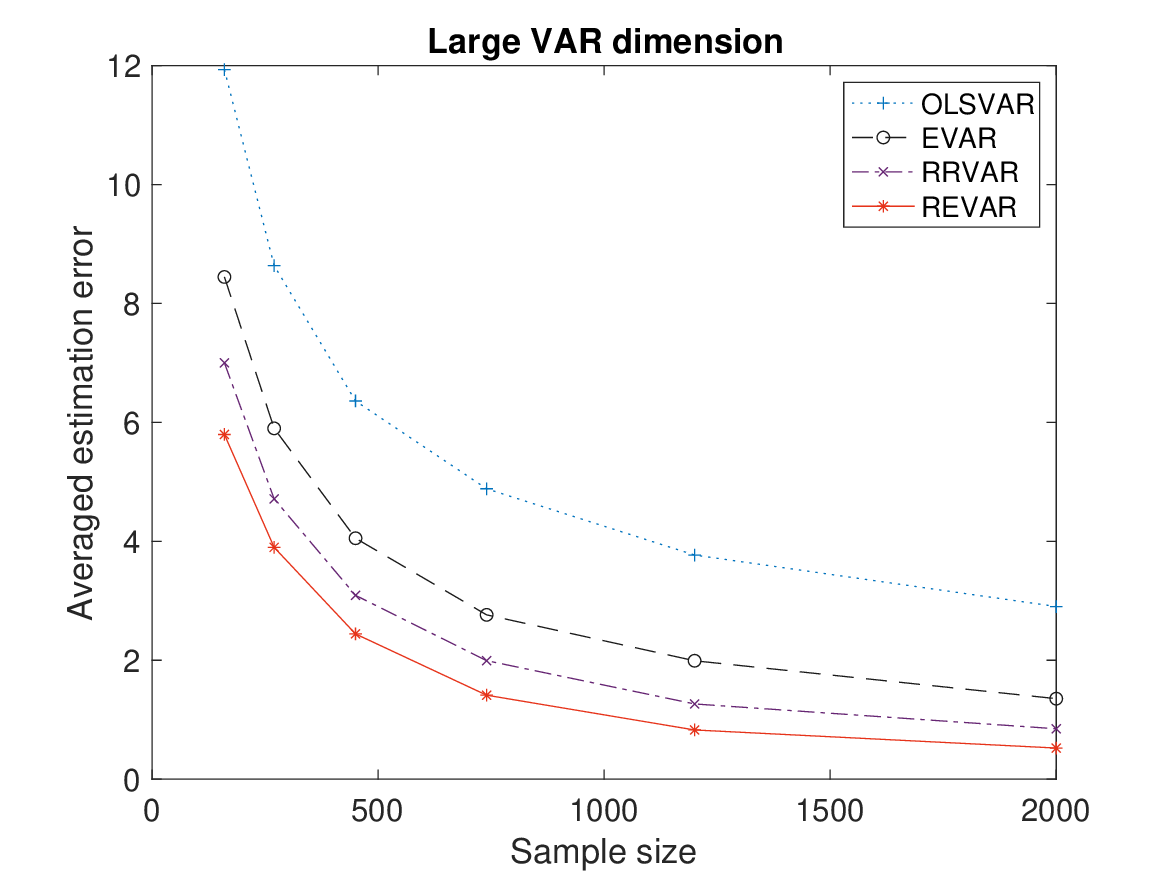}
                \caption{ $(d, u, p, q)=(10, 20, 1, 40)$}
        \end{subfigure}%
      \hspace{-.5cm}  
           \begin{subfigure}[b]{0.5\textwidth}
                \includegraphics[width=\textwidth,height=4.5cm]{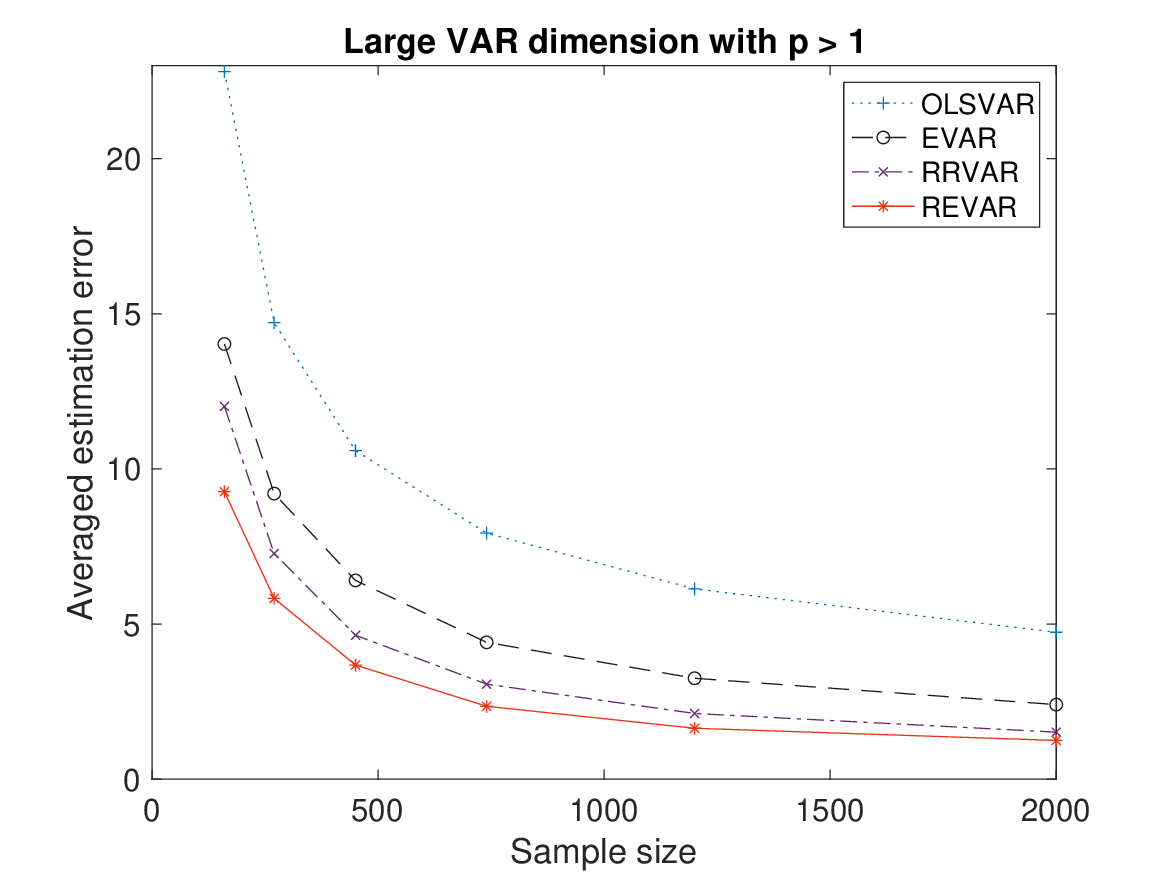}
             \caption{ $(d, u, p, q)=(10, 20, 2, 40)$}
        \end{subfigure}
     \vspace{.3cm}
      \caption{ Average estimation errors of higher-dimensional VAR models with different combinations of $(d, u, p, q)$  against sample size.  Refer to Table \ref{tb:1.1} for details }
\label{fig2}
   \end{figure}

 To further investigate the performance of the proposed REVAR model, particularly in higher-dimensional cases, we demonstrate its effectiveness in four additional examples involving medium and large VARs. Table \ref{tb:1.1} compares the total number of parameters for each model with different combinations of $(d, u, p, q)$. Similar to Figure \ref{fig1}, Figure \ref{fig2} illustrates the impact of envelope dimension and rank on the comparative performances of each approach. The top row of Figure \ref{fig2} graphically presents the comparison of the four VAR models
 for medium-sized VARs with $p=1$ and $p>1$, while the bottom row shows the comparison  
for relatively large VARs. All simulation results demonstrate the superiority of our  REVAR model with significant improvements over other VAR models. Due to computational costs,    
the 1D algorithm is used for the simulation studies of medium and large VARs, as it offers faster computation compared to the FG algorithm.  

\begin{table}[H]
        \small
	\renewcommand{\arraystretch}{.6}
	\caption{Total number of parameters (NOP) in each VAR model  correspond to Figure \ref{fig2}}
	\begin{center}
		\begin{tabular}{c c c c c c} \hline
		\centering
			      & &Medium VAR  &Medium VAR with $p\! > \!1$  &Large VAR  &Large VAR with $p\! >\! 1$\\
			        &  $(d, u, p, q)$ &  $(5, 10, 1, 20)$ &$(5, 10, 2, 20)$ & $(10, 20, 1, 40)$ &$(10, 20, 2, 40)$\\
			      \hline
				
			      &OLSVAR &610  &1010    &2420   &4020\\
				  &RRVAR     &385 &485    &1520    &1920\\
                  &EVAR    &410 &610    &1620    &2420\\
				  &REVAR    &335  &435    &1320        &1720\\

				         \hline
		\end{tabular}
	\end{center}
	\label{tb:1.1}
\end{table}

\begin{figure}[H]
 %\vspace{-1.cm}
   \centering
 \begin{subfigure}[b]{0.5\textwidth}
    \includegraphics[width=\textwidth,height=4.4cm]{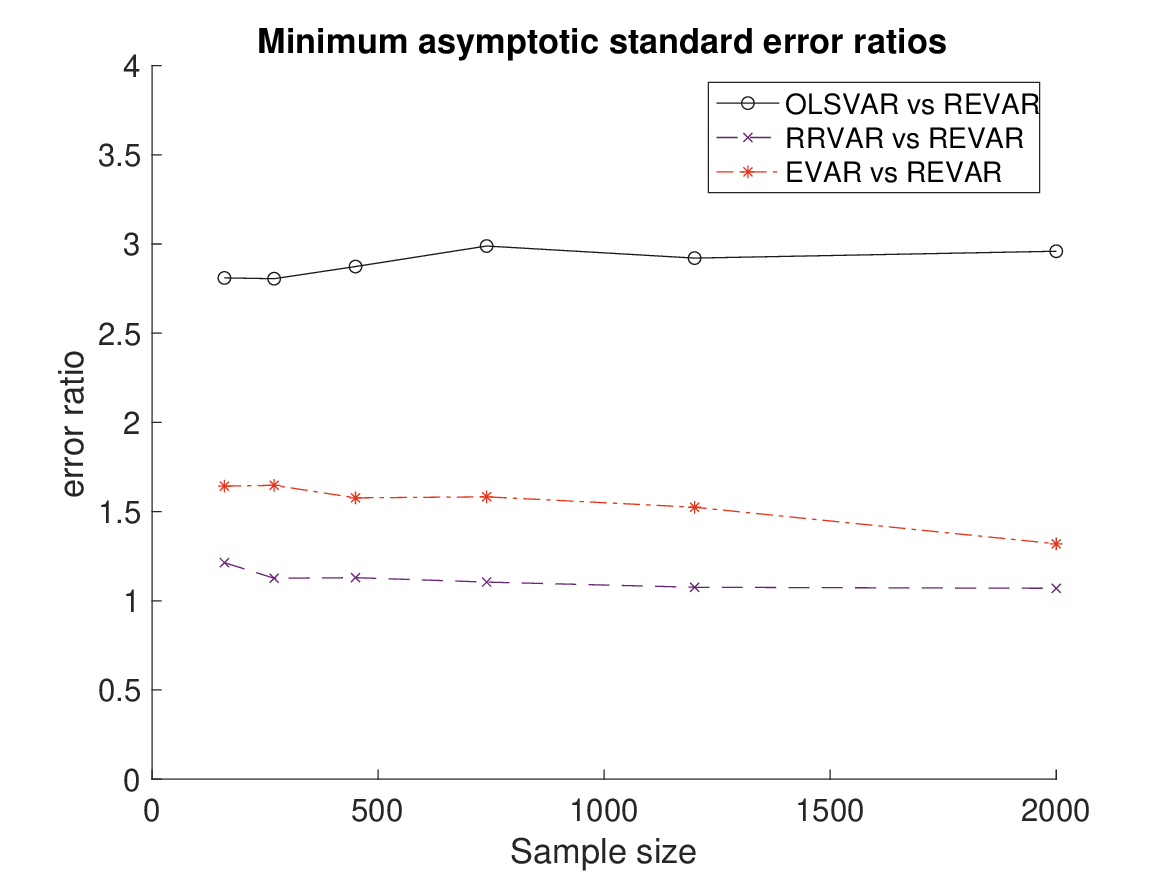}
               \caption{ $(d, u, p, q)=(5, 10, 1, 20)$}
        \end{subfigure}%
         \hspace{-.5cm}
           \begin{subfigure}[b]{0.5\textwidth}
    \includegraphics[width=\textwidth,height=4.4cm]{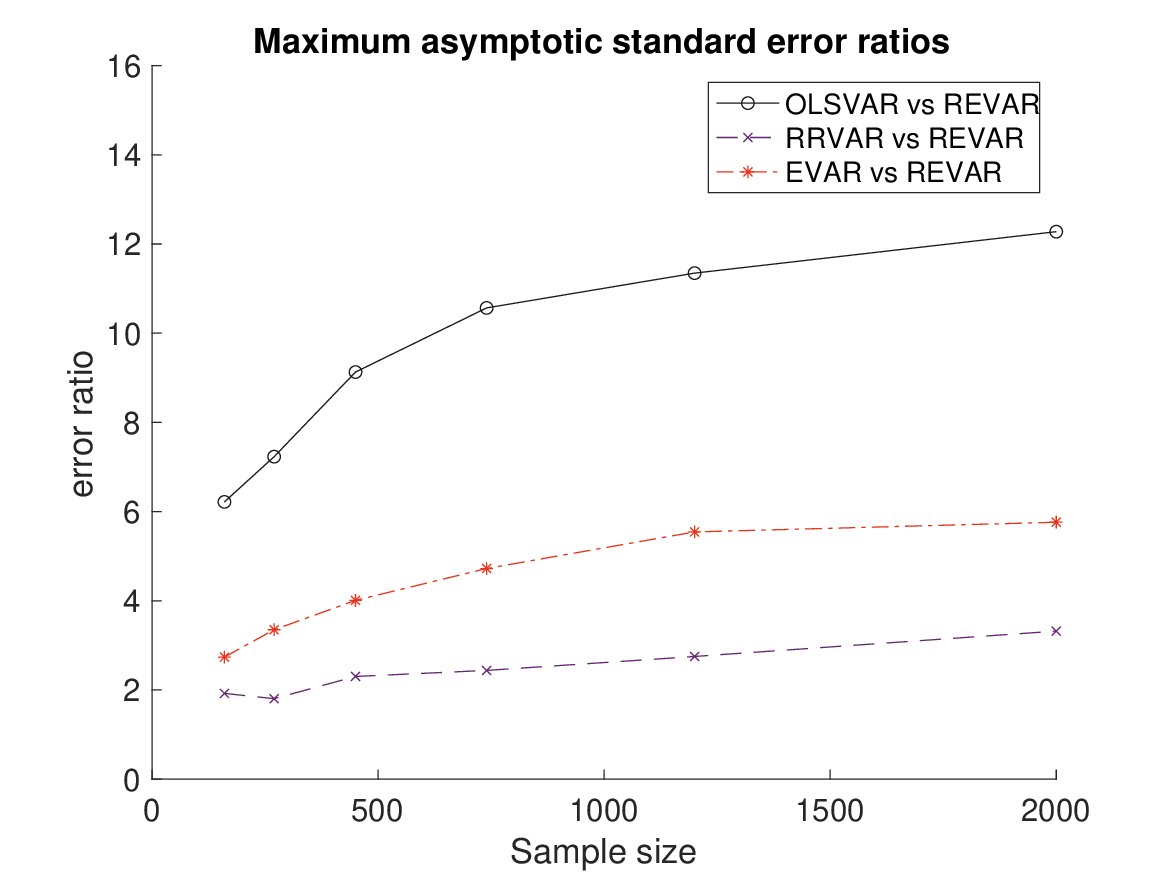}
               \caption{ $(d, u, p, q)=(5, 10, 1, 20)$}
        \end{subfigure}
    %\vspace{-.1cm}
        \centering
 \begin{subfigure}[b]{0.5\textwidth}
    \includegraphics[width=\textwidth,height=4.4cm]{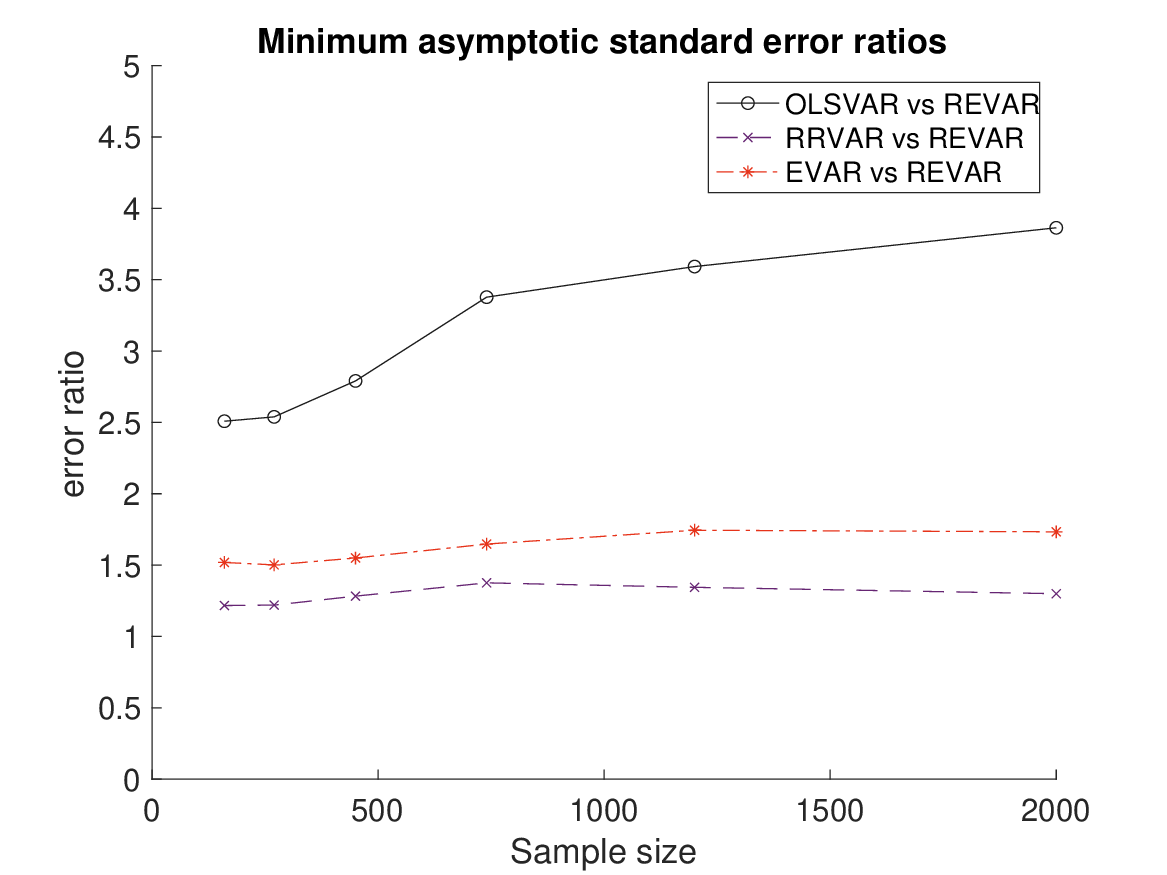}
               \caption{$(d, u, p, q)=(10, 20, 1, 40)$}
        \end{subfigure}%
      \hspace{-.5cm}   
           \begin{subfigure}[b]{0.5\textwidth}
  \includegraphics[width=\textwidth,height=4.4cm]{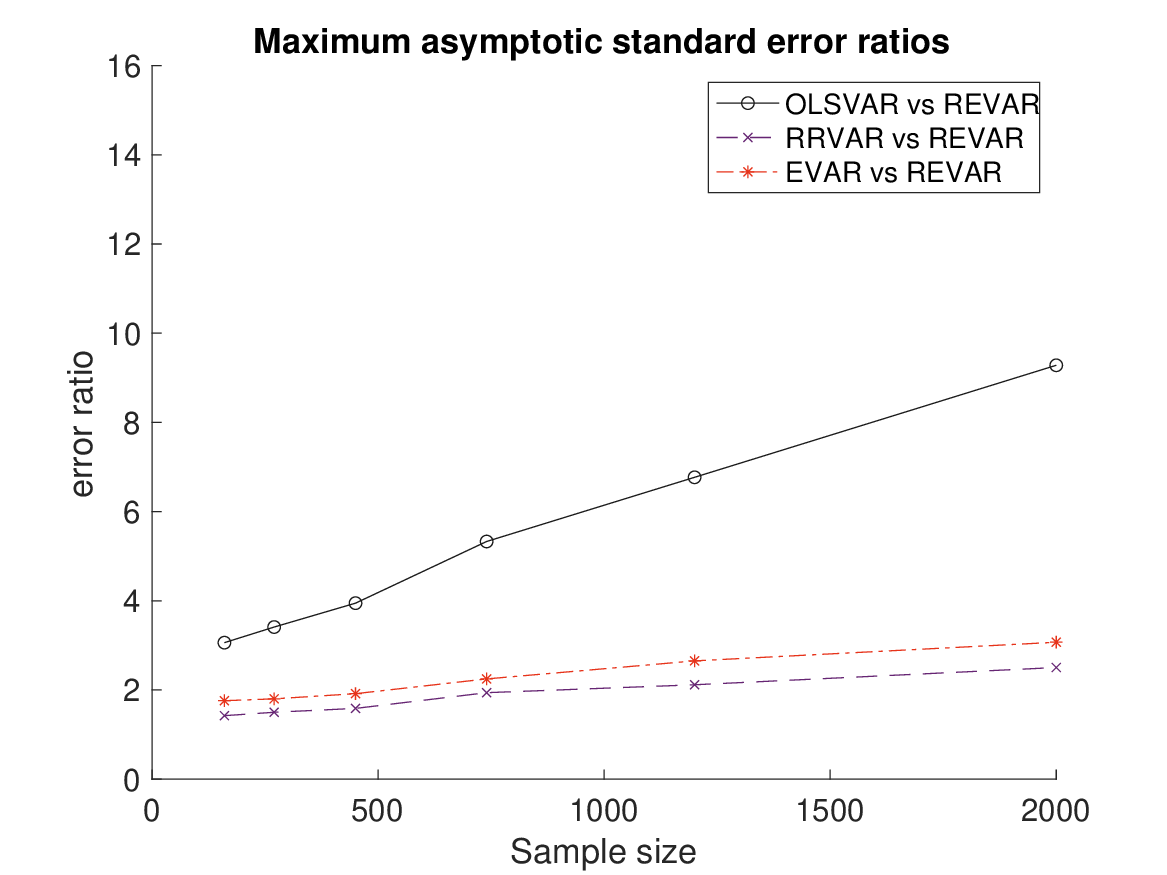}
               \caption{ $(d, u, p, q)=(10, 20, 1, 40)$}
        \end{subfigure}
     \vspace{.3cm}
     \caption{Minimum (left panels) and maximum (right panels) asymptotic standard error ratios  of coefficient estimates  for each VAR model with respect to  the proposed REVAR model}
\label{fig2.1}
\end{figure}

Figure \ref{fig2.1} summarizes the minimum and maximum asymptotic standard error ratios of the estimated coefficients for each of the OLSVAR, EVAR, and RVAR models in relation to the proposed REVAR model. The top row of  Figure \ref{fig2.1}  displays these ratios versus the sample size when  $(d, u, p, q)=(5, 10, 1, 20)$,  while the bottom row   shows  the ratios   for   $(d, u, p, q)=(10, 20, 1, 40)$.  
All ratios in Figure \ref{fig2.1} are greater than one,   indicating that the REVAR model achieves the highest efficiency gains by significantly reducing the standard errors of the estimated coefficient matrix compared to the three comparative VAR models.

%\vspace{-.3in}

%\vspace{-.2in}

\vspace{-.2in}
\subsection{Simulation Studies with Non-normal Errors}\label{Simulation_nonnormal}
\begin{figure}[H]
 %\vspace{-1.cm}
   \centering
 \begin{subfigure}[b]{0.5\textwidth}
  \includegraphics[width=\textwidth,height=4.6cm]{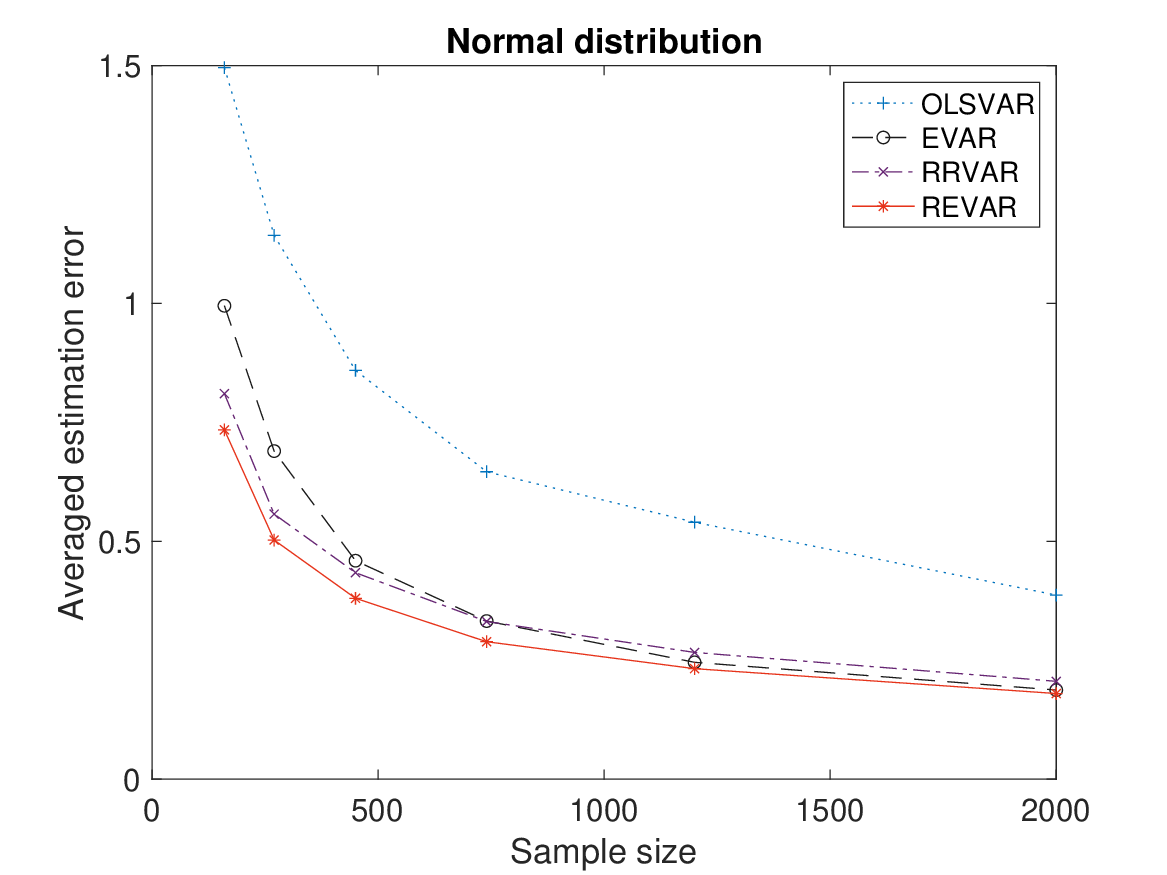}
                \caption{ $(d, u, p, q)=(3, 4, 1, 7)$}
        \end{subfigure}%
         \hspace{-.6cm}
        %add desired spacing between images, e. g. ~, \quad, \qquad etc.
   \begin{subfigure}[b]{0.5\textwidth}
         \includegraphics[width=\textwidth,height=4.6cm]{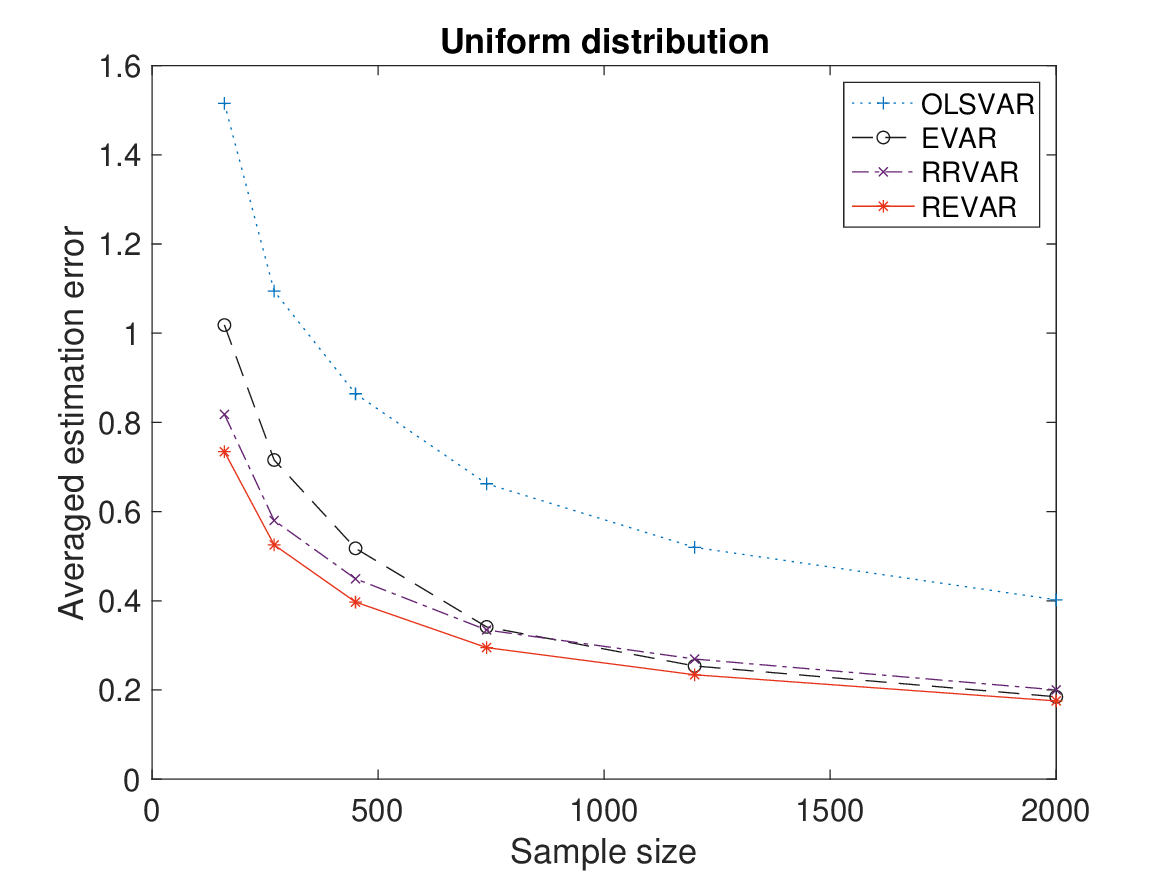}
                \caption{$(d, u, p, q)=(3, 4, 1, 7)$ }
        \end{subfigure}
    %\vspace{-.1cm}
        \centering
 \begin{subfigure}[b]{0.5\textwidth}
                \includegraphics[width=\textwidth,height=4.6cm]{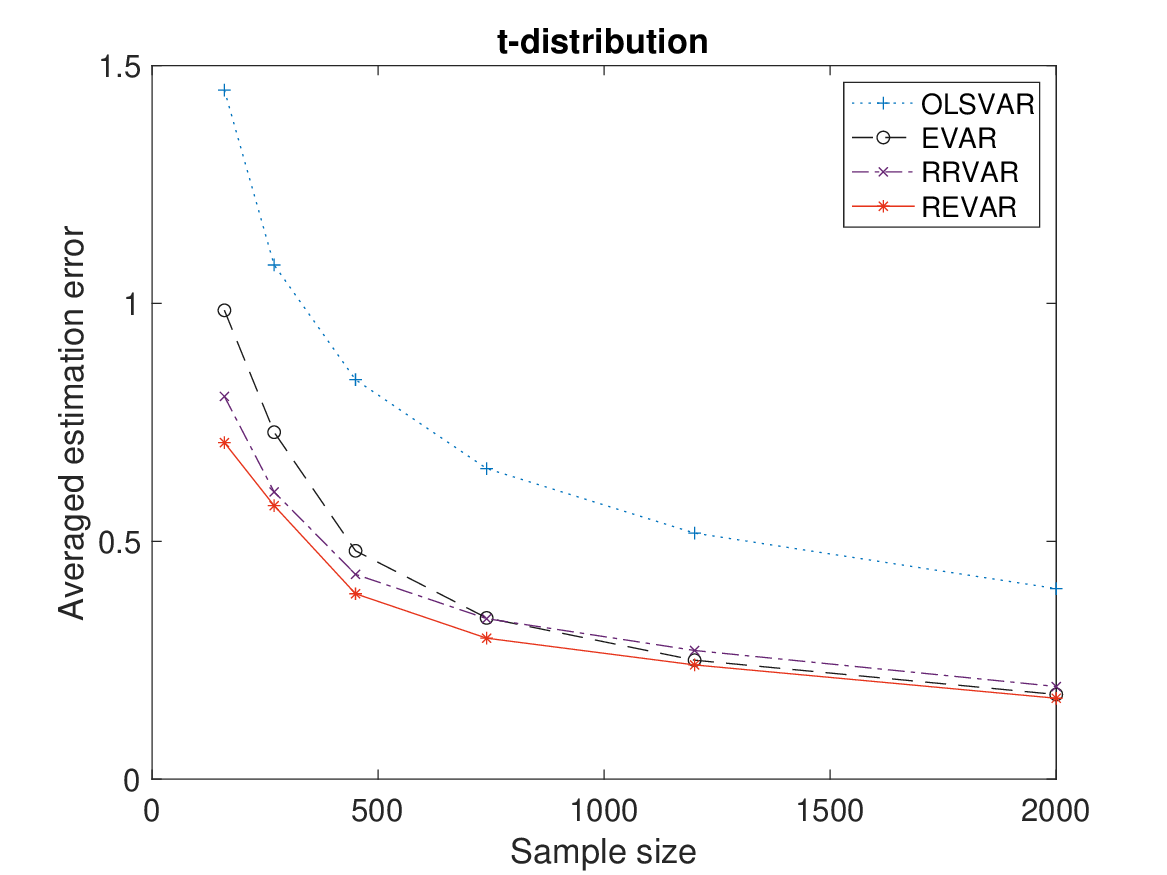}
                \caption{ $(d, u, p, q)=(3, 4, 1, 7)$}
        \end{subfigure}%
      \hspace{-.6cm}  %add desired spacing between images, e. g. ~, \quad, \qquad etc.
           \begin{subfigure}[b]{0.5\textwidth}
                \includegraphics[width=\textwidth,height=4.6cm]{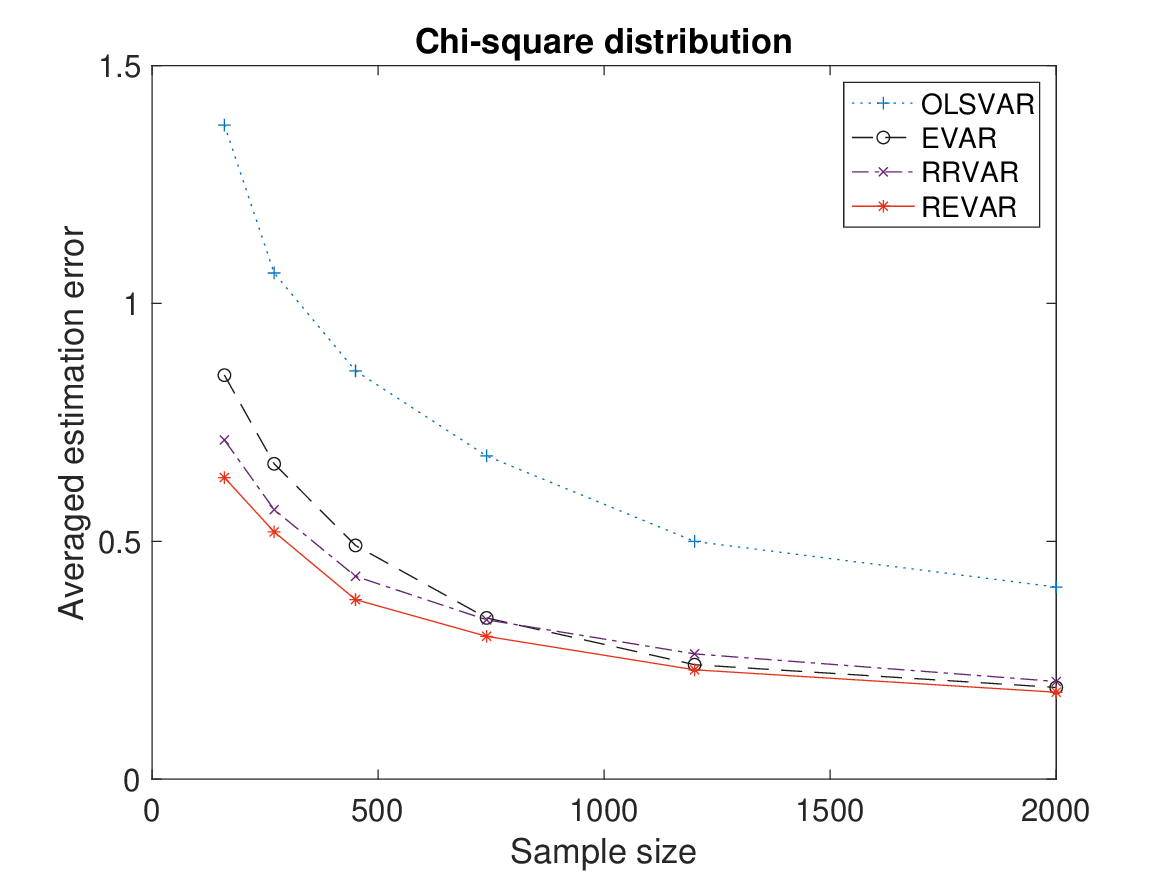}
               \caption{$(d, u, p, q)=(3, 4, 1, 7)$ }
        \end{subfigure}
     \vspace{.3cm}
    \caption{Average estimation errors  for      $(d, u, p, q)=(3, 4, 1, 7)$   with Normal, Uniform, t-Student, and Chi-square $\boldsymbol\varepsilon_t$ distributions, plotted against sample size. Refer to  Table \ref{tb:1} for details}
       \label{fig5}
   \end{figure}

\vspace{-.05in}
In this subsection, simulations are conducted with non-normal errors using  $\bvarepsilon_t = \bSigma^{1/2}\boldsymbol\Upsilon_t$, where $\boldsymbol\Upsilon_t$ is a vector of i.i.d random variables with a  mean of  $\mathbf{0}$ and a covariance matrix of  $\mathbf{I}_q$. We considered four different distributions, i.e., normal, uniform, $t_6$, and $\chi^2_6$ distributions.  
 Figure \ref{fig5} compares the relative performance of each method across these distributions,  in terms of the impact of the envelope dimension and rank.
 Our experiments demonstrated that the REVAR estimator consistently has the lowest average estimation error compared to the other models across all four distributions.   Similar results were obtained for higher-dimensional VAR models, which are presented in Figures S1 and S2 % \ref{fig8.0} and \ref{fig8.0.1} 
 in Supplement S8.1. 
 All the minimum and maximum asymptotic standard error ratios of each VAR model relative to our proposed REVAR model exceed one, indicating that the REVAR model consistently achieves higher efficiency gains  (see Figures  S3,  S4 and S5 in Supplement S8.1).  %(see Figures \ref{fig8.uf}, \ref{fig8.ch} and \ref{fig8.t6} in Supplement S8.1).

\vspace{-.35in}
\subsection{\!\! Simulation Studies with Martingale Difference  Errors}\label{Simulation_Martingale}
 
 In this subsection, we adopt a martingale difference error structure for the analysis.  
Let $\{\bvarepsilon_{t}\}$ be a $q$-dimensional martingale difference sequence  with respect to the increasing sequence of  $\sigma$-fields $\{\mathcal{F}_{t} $ \},  
 where $\bvarepsilon_{t}$ are $\mathcal{F}_{t}$-measurable and $E(\bvarepsilon_{t}|\mathcal{F}_{t}) = 0$ for all $1\leq t \leq T$. We also introduce a martingale sequence  $\{\bzeta_{t}, \sigma(\bzeta_{t}) \}$ for each $t$,  
and define  the error term  as $\bvarepsilon_{t} = \bzeta_{t+1} - \bzeta_{t}$.   
To simulate the error structure, we generate
$\bzeta_{1}$ from $\mathcal{N}(0,\bSigma)$ distribution and subsequently generate the rest of the sequence from the conditional distribution $\bzeta_{t+1}|\bzeta_{t} \sim \mathcal{N}(\bzeta_{t}, \bSigma)$. This simulation method has been used in previous studies (Zhao et al., 2011; Zhou and Lin, 2013). This simulation method has also been applied to non-normal cases. 

Furthermore, we adopt a stochastic volatility martingale difference sequence model to capture the time-varying volatility in the time series data.  The error terms are modeled as 
\begin{equation*}
\bvarepsilon_t= \mathbf{e}_t  \exp(\boldsymbol\sigma_t) \quad  \text{with}  \quad  \boldsymbol\sigma_t = 0.25\boldsymbol\sigma_{t-1} + 0.05 \mathbf{u}_t,  \quad  \mathbf{e}_t  \overset{\mathrm{i.i.d}}{\sim} \mathcal{N}(\mathbf{0}, \boldsymbol\Sigma),
\quad \mathbf{u}_t \overset{\mathrm{i.i.d}}{\sim}  \mathcal{N}(\mathbf{0}, \mathbf{V}_{u}),
\end{equation*}
 where  
 $\mathbf{V}_{u} = (\nu_{u,kl}) \in \mathbb{R}^{p \times p}$ with 
 $\nu_{u,kl} = 0.9^{|k - l|}$.  
 This simulation method has been  previously utilized  by Chang et al. (2022), and  Escanciano and Velasco (2006) for the univariate and multivariate  martingale difference
hypothesis testing problems, respectively.

\begin{figure}[H] 
 %\vspace{-1.cm}
   \centering
 \begin{subfigure}[b]{0.5\textwidth}
  \includegraphics[width=\textwidth,height=4.6cm]{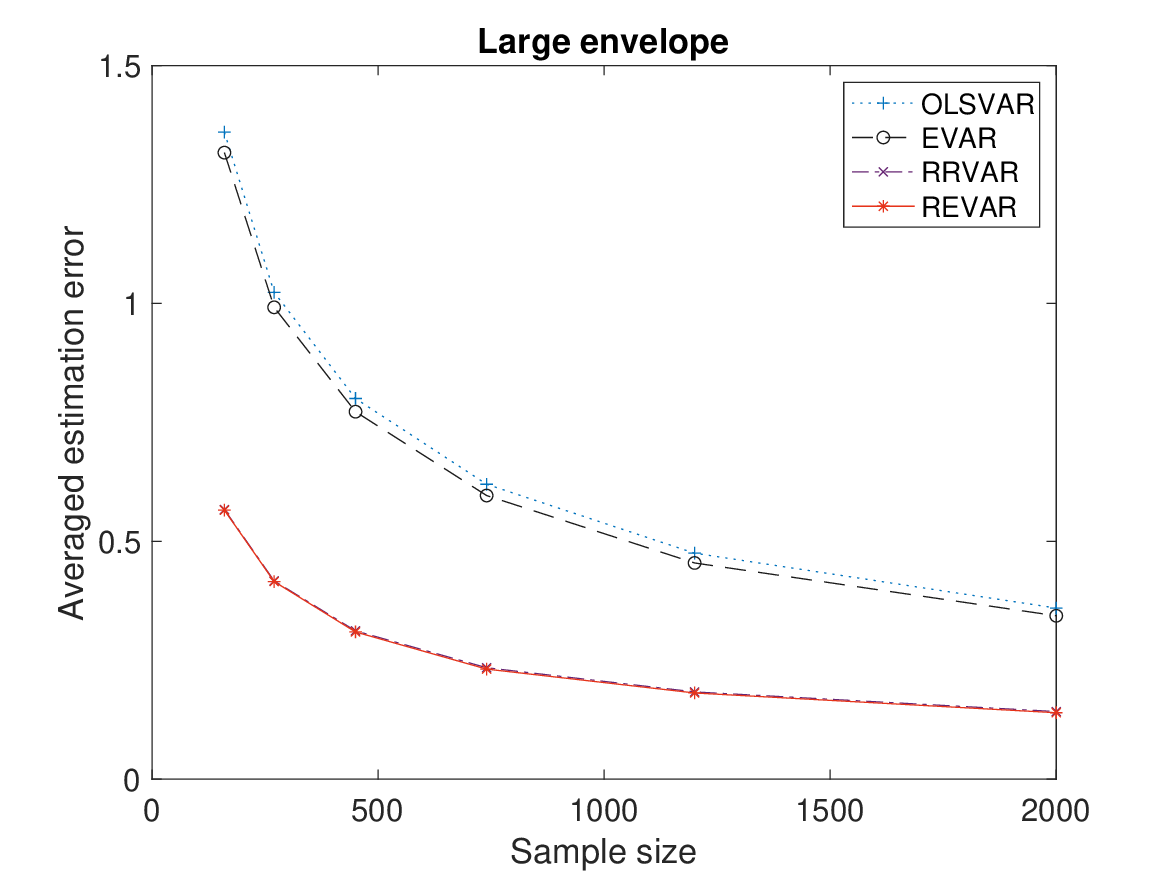}
                \caption{ $(d, u, p, q)=(2, 6, 1, 7)$}
        \end{subfigure}%
         \hspace{-.6cm}
        %add desired spacing between images, e. g. ~, \quad, \qquad etc.
   \begin{subfigure}[b]{0.5\textwidth}
         \includegraphics[width=\textwidth,height=4.6cm]{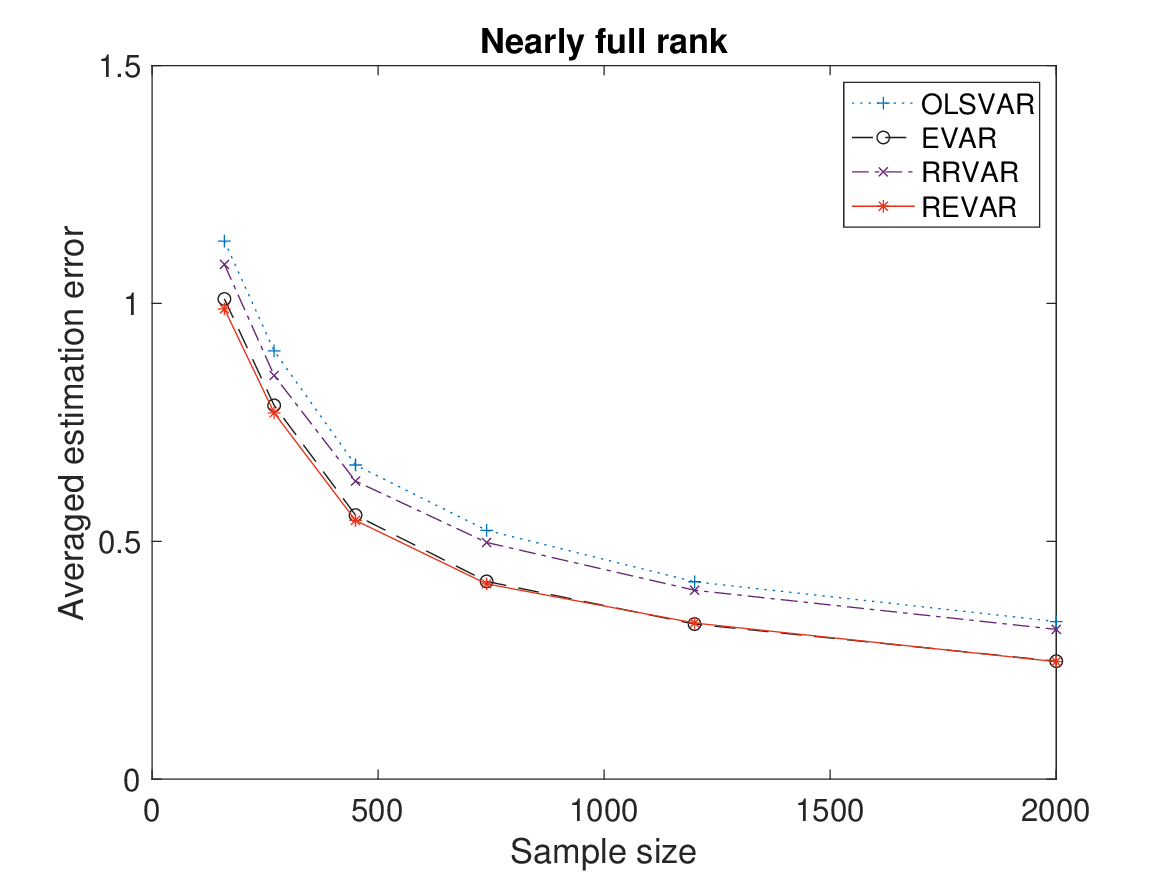}
                \caption{$(d, u, p, q)=(5, 6, 1, 7)$ }
        \end{subfigure}
    %\vspace{-.1cm}
        \centering
 \begin{subfigure}[b]{0.5\textwidth}
            \includegraphics[width=\textwidth,height=4.6cm]{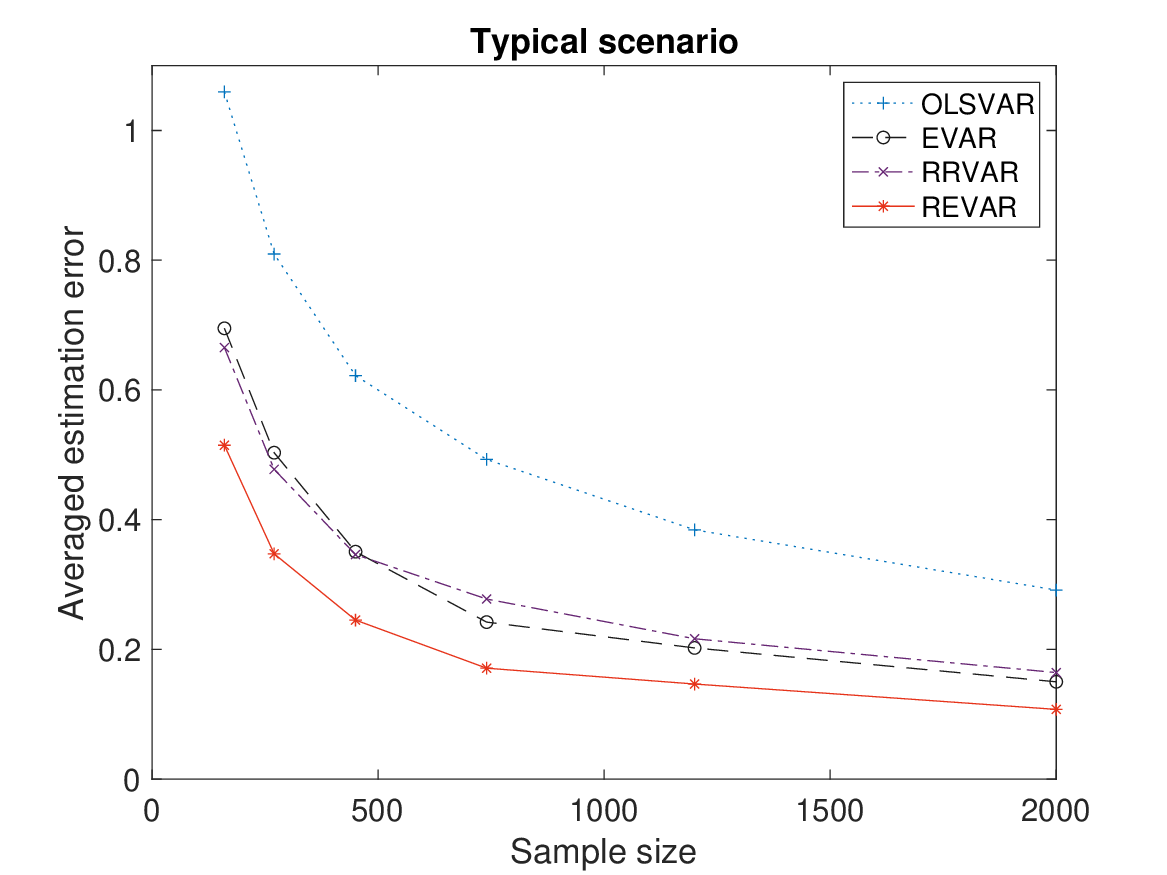}
                \caption{ $(d, u, p, q)=(3, 4, 1, 7)$}
        \end{subfigure}%
      \hspace{-.6cm} 
           \begin{subfigure}[b]{0.5\textwidth}
            \includegraphics[width=\textwidth,height=4.6cm]{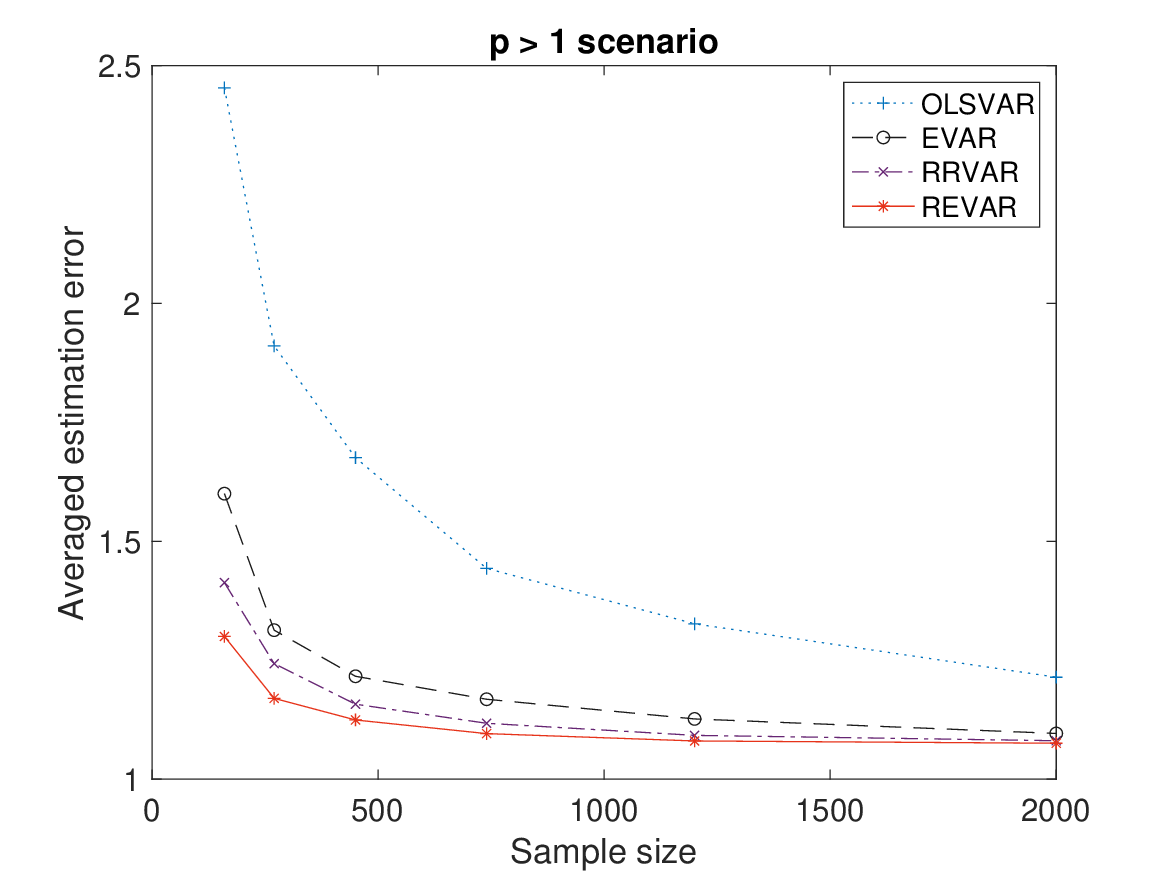}
               \caption{$(d, u, p, q)=(3, 4, 2, 7)$ }
        \end{subfigure}
     \vspace{.3cm}
    \caption{Average estimation errors of moderate-dimensional VAR models with different     $(d, u, p, q)$     for stochastic volatility  martingale difference sequence (SV-MDS) errors against sample size}
\label{fig7}
   \end{figure}

Figure S6  %\ref{fig6} 
displays the impact of envelope dimension and rank  when errors are generated from 
martingale difference sequences, while Figure \ref{fig7}  presents the results for errors generated from stochastic volatility martingale difference sequences. Across both figures, the simulation results consistently demonstrate the superior performance of our proposed REVAR model, revealing significant improvements over other VAR models. Similar findings were observed for higher-dimensional VAR models, as shown in Figures S7 and S8 % \ref{fig8} and \ref{fig9} 
in Supplement S8.2. Furthermore, all the minimum and maximum asymptotic standard error ratios of each VAR model relative to our proposed REVAR model were found to be greater than one, indicating the consistent efficiency gains achieved by the REVAR model (refer to Figures S9 and S10  % \ref{fig8.r1} and \ref{fig8.r2}
 in Supplement S8.2)

% \vspace{-.1in}

   %%%%%%%%%%%%

We conduct a pseudo-real-time forecasting exercise  to calculate $h$-step ahead forecasts ($\widehat{y}_{t+h|t}$).    Forecast accuracy is obtained by calculating 
%The accuracy of these forecasts is  evaluated using 
 the root mean square forecast error (RMSFE): 
\begin{equation}\label{rmsfe}
    \textup{RMSFE}_h  = \sqrt{\frac{1}{T-T_0-\mathcal{H}+1}\sum_{t = T_0+\mathcal{H}-h}^{T-h}(\widehat{y}_{t+h|t} - y_{t+h})^2}, ~~~~h=1,\dots, \mathcal{H},
\end{equation}
where $T_0$ and $T$ indicate the start and the
end of the evaluation sample, respectively, and $\mathcal{H}$ is the maximum forecast horizon of interest. 
Our forecasting simulations encompass various error distributions, such as normal, non-normal, and martingale difference errors with stochastic volatility (SV). The evaluation sample consists of the last 25\%   of the data. 

Table S2  %\ref{tb.3}  
and Table S3  %\ref{tb.4} 
in Supplement S8.3 present a comprehensive summary of the root mean square forecast error (RMSFE$_h$) %for $h$-step ahead
for different forecast horizons ($h=1,2,3,4$), averaged over the evaluation samples across various error distributions and different DGP scenarios,  with a sample size of $T=700$. The results consistently demonstrate that our REVAR model outperforms other VAR models in terms of forecast accuracy. 
Furthermore, all average asymptotic standard error ratios relative to the REVAR model exceed one, indicating greater efficiency gains and improved accuracy in coefficient estimation compared to the comparative VAR models.

Table \ref{tb.2.1} summarizes the asymptotic chi-squared test results  
for rank selection, envelope dimension, and lag order selections using the BIC criterion at a $0.05$ significance level. We performed the simultaneous selection of $d$ and $u$.    The REVAR model with dimensions $(d, u, p, q) = (3, 5, 1, 7)$  and $(2, 4, 2, 8)$ are utilized,  with a total  NOP in REVAR model of $55$  and $72$, respectively.   The table reports   the   percentages  of correctly identified rank $d$ and envelope VAR dimension $u$ and lag order $p$  
for different sample sizes.  
Most of the other combinations of $(d, u, p, q)$ employed in the simulation studies yielded similar results, although in some cases, the selection of $u$ and $d$ was less accurate and resulted in higher percentages of overestimation.
 Consistent with the findings of Forzani and Su (2021), our results indicate that when the information criteria fail to select the true dimensions ($u$ and $d$), they tend to overestimate these dimensions. Although this overestimation leads to a loss of efficiency, it does not introduce bias into the estimation process.
 Conversely, underestimation of $d$ and $u$ can indeed lead to biased outcomes. Nevertheless,  overestimation of $d$ and $u$ is typically not a major concern (Cook et al., 2015).

\begin{table}[H]
        \footnotesize
	\renewcommand{\arraystretch}{.7}
	%\caption{Predicted model for GDP and its decomposition using bootstrap approach ($d, u, p, q$) = (3, 4, 1, 8)}
	\caption{Percentage selection of the true dimensions ($d$ and $u$) and lag order ($p$)}
	\begin{center}
	%	\begin{tabular}{lllllrrrr} \hline
  		\begin{tabular}{cccccccc} \hline
		\centering
  & \multicolumn{3}{c}\text{$(d, u, p, q) = (3, 5, 1, 7)$}  &\multicolumn{3}{c}\text{$(d, u, p, q) = (2, 4, 2, 8)$} & \\ \cmidrule(lr){3-5} \cmidrule(lr){6-8}
 \vspace{-.03in}			     &    &\multicolumn{1}{c}{$p$ Selection} &\multicolumn{1}{c}{$d$ Selection} &\multicolumn{1}{c}{$u$ Selection} &\multicolumn{1}{c}{$p$ Selection} &\multicolumn{1}{c}{$d$ Selection} &\multicolumn{1}{c}{$u$ Selection} \\ 
        & T &BIC & Chi-squared  test &BIC &BIC & Chi-squared test &BIC \\
        \hline
			      &  240  &100\% &93\%   &88\%  &100\% &83\%   &54\%   \\
				  &  450  &100\%  &95\%    &92\%  &100\%  &94\%    &90\%       \\
                    &  700  &100\%  &97\%   &99\%  &100\%  &94\%   &99\%      \\
				  &  950  &100\%  &98\%   &100\%  &100\%  &93\%   &98\%      \\
                    & 1200  &100\%  &98\%   &100\%  &100\%  &94\%   &99\%       \\
				         \hline
		\end{tabular}
	\end{center}
	\label{tb.2.1}
\end{table}

\vspace{-.6in}

\section{Real Data Analysis}\label{Real data analysis}
\vspace{-.1in}
  
We analyzed four quarterly macroeconomic datasets obtained from the Federal Reserve Economic Quarterly Data (FRED-QD) website, with $T = 244$ observations from 1959Q1 to 2019Q4. The first dataset, NIPA, consists of $q = 8$ macroeconomic variables from the National Income and Product Accounts, including real Gross Domestic Product (GDP) and its components. 
The second dataset consists of $q=11$ variables related to Industrial Production, the third dataset includes $q=20$ variables related to Price, and the fourth dataset encompasses $q=8$ variables related to Money and Credit.
Table S6  %\ref{t:fc1} 
 in Supplement S9 provides a detailed overview of the variables in each dataset.     To ensure stationarity, appropriate transformations were applied to each variable group,  such as taking the first difference of the logarithmic series for the $8$ NIPA variables.  
Further details regarding variable descriptions and transformations can be found in McCracken and Ng (2020).  
  
  The rank ($\widehat{d}$)  is determined
  using the chi-squared  test  (Section \ref{Rank ($d$) selection}) at a significance level of $0.01$. The lag order ($\widehat{p}$) and the envelope dimension ($\widehat{u}$) are selected based on the BIC criterion as described in Sections \ref{Lag order ($p$) selection}, and   \ref{Envelope VAR dimension ($u$) selection}, respectively.  
  Pseudo-real-time forecasting experiments were conducted using the evaluation sample (2005Q1-2019Q4) employing the stationary bootstrap scheme proposed by Politis and Romano (1994) with 100 bootstrap samples. The forecasting performance of our proposed REVAR model was evaluated using the $\textup{RMSFE}_h$  metric for different forecast horizons $h$ and compared with other VAR models.
 
%We conducted pseudo-real-time forecasting experiments using an expanding window scheme. The process begins with a specific initial fraction of the full sample (1959Q1-2004Q4) and recursively calculated forecasts over the evaluation sample (2005Q1-2019Q4). These forecasts are computed using the stationary bootstrap scheme proposed by Politis and Romano (1994) with $100$ bootstrap samples. We assessed the forecasting performance of our proposed REVAR model using the $\textup{RMSFE}_h$  for different forecast horizons ($h$) and compared it with other VAR models.

Table \ref{tb.5} and Table \ref{tb.6} present  the RMSFE$_h$ values  ($h = 1, 2, 3, 4$)  for the NIPA and Price datasets, respectively. Supplementary Tables S4 and S5  %\ref{tb.7} and \ref{tb.8} 
in Supplement S9 provide the same results for the Money and Credit, and the  Industrial Production datasets, respectively. 
The results consistently demonstrate the superior forecasting performance of our proposed REVAR model compared to the three competitive VAR models.  
Moreover, the average asymptotic standard error ratios ($r_{\tiny\textup{avg.}}$) of the OLSVAR, EVAR, and RRVAR  models relative to our proposed REVAR model are consistently greater than one, which indicates significant efficiency gains achieved by the REVAR model.
 
%%%%%%%%%%%%Final tables

\begin{table}[H]
        \small
	\renewcommand{\arraystretch}{.7}
	%\caption{Predicted model for GDP and its decomposition using bootstrap approach ($d, u, p, q$) = (3, 4, 1, 8)}
	\caption{Pseudo-real-time forecasting performance with bootstrap for  the NIPA dataset (1959Q1-2019Q4)  calculated over   the evaluation sample period 
 %using an expanding window scheme, evaluated
 from  2005Q1 to 2019Q4}%   for the years 1959Q1 to 2019Q4}
 \vspace{-.05in}
	\begin{center}
	%	\begin{tabular}{lllllrrrr} \hline
   		\begin{tabular}{ccccccccc} 
     \hline
		\centering
    &  &   \multicolumn{5}{c}\text{RMSFE$_h$} & \\ \cline{6-9}
			      &$(\widehat{d}, \widehat{u}, \widehat{p}, q)$   &Model(M) & NOP &$r_{\tiny\textup{avg.}}$   &h=1 &  h=2 & h=3 & h=4 \\ \hline
			      &    &OLSVAR &100  &1.2912      &0.059709    &0.059433    &0.059647    &0.059671\\
				  & (3, 4, 1, 8)      &EVAR &68  &1.0854      &0.059133     &0.05888    &0.059074     &0.05909\\
                  &               &RRVAR &75 &1.0029 &0.059278    &0.058971    &0.059047    &0.059086 \\
				  &              &REVAR  &63   &$~\_$       &0.058950    &0.058666     &0.058850    &0.058788 \\
				         \hline
		\end{tabular}
	\end{center}
	\label{tb.5}
\end{table}

\vspace{-.25in}
 
\begin{table}[H]
        \small
	\renewcommand{\arraystretch}{.7}
	%\caption{Predicted model for GDP and its decomposition using bootstrap approach ($d, u, p, q$) = (3, 4, 1, 8)}
	\caption{Pseudo-real-time forecasting performance with bootstrap for the Price dataset   (1959Q1-2019Q4) calculated over the evaluation sample period 
 from  2005Q1 to 2019Q4}
 \vspace{-.05in}
	\begin{center}
	%	\begin{tabular}{lllllrrrr} \hline
   		\begin{tabular}{ccccccccc} 
     \hline
		\centering
    &  &   \multicolumn{5}{c}\text{RMSFE$_h$} & \\ \cline{6-9}
			      &$(\widehat{d}, \widehat{u}, \widehat{p}, q)$   &Model(M) & NOP &$r_{\tiny\textup{avg.}}$   &h=1 &  h=2 & h=3 & h=4 \\ \hline
			      &    &OLSVAR &610  &1.2236      &0.13261     &0.13114     &0.13214     &0.13187\\
				  & (13, 14, 1, 20)      &EVAR &490  &1.0203 &0.13194     &0.13065     &0.13156     &0.13112\\
                  &               &RRVAR &561 &1.1201 &0.13228     &0.13105     &0.13182     &0.13157 \\
				  &              &REVAR  &483   &$~\_$       &0.13184     &0.13063      &0.13150     &0.13101 \\
				         \hline
		\end{tabular}
	\end{center}
	\label{tb.6}
\end{table}
\vspace{-.25in}

In order to evaluate the effectiveness of the proposed REVAR model with a larger number of variables,   Table \ref{tb.9} summarizes the RMSFE$_h$ values  ($h = 1, 2, 3, 4$)  using the first 43  variables from the FRED-QD dataset. These results not only show the superior performance of the REVAR model over other VAR models but also highlight its significant efficiency gains and forecasting improvements compared to smaller models.  

\begin{table}[H]
        \small
	\renewcommand{\arraystretch}{.7}
	%\caption{Predicted model for GDP and its decomposition using bootstrap approach ($d, u, p, q$) = (3, 4, 1, 8)}
	\caption{Pseudo-real-time forecasting performance with bootstrap for  the initial 43   out of 47  macroeconomic variables in Table S6  %\ref{t:fc1} 
	(1959Q1-2019Q4),  %calculated over   the evaluation sample period
 evaluated from  2005Q1 to 2019Q4}  %using an expanding window scheme}%   for the years 1959Q1 to 2019Q4}
\vspace{-.05in}	
 \begin{center}
	%	\begin{tabular}{lllllrrrr} \hline
   		\begin{tabular}{ccccccccc} 
     \hline
		\centering
    &  &   \multicolumn{5}{c}\text{RMSFE$_h$} & \\ \cline{6-9}
			      &$(\widehat{d}, \widehat{u}, \widehat{p}, q)$   &Model(M) & NOP &$r_{\tiny\textup{avg.}}$   &h=1 &  h=2 & h=3 & h=4 \\ \hline
			      &    &OLSVAR &2795  &1.4202      &0.21493     &0.21371     &0.21381     &0.21396\\
				  & (18, 19, 1, 43)      &EVAR &1763 &1.0406  &0.20709     &0.20664     &0.20735     &0.20668    \\
                  &               &RRVAR &2170 &1.0594 &0.21003     &0.20956     &0.20962     &0.20925 \\
				  &              &REVAR  &1738   &$~\_$       &0.20644     &0.20615     &0.20681     &0.20631 \\
      \hline
		\end{tabular}
	\end{center}
	\label{tb.9}
\end{table}

%%%%%%%%%%%%%%%%%%%%%%%%%%%%%%%%%%%%%%%%

\vspace{-.35in}
\section{Discussion}\label{Discussion}
 \vspace{-.15in}

In this paper, we proposed the reduced-rank envelope VAR (REVAR) model, a parsimonious vector autoregressive model that incorporates the concept of envelopes into the reduced-rank VAR framework. 
The proposed REVAR approach offers a powerful approach to detect relevant and irrelevant information in VAR models and significantly improve estimation accuracy and efficiency by removing the immaterial information and variations from the VAR process. We derived maximum likelihood (ML) estimators for the REVAR model and investigated their asymptotic properties 
under various error assumptions, including normality, non-normality, and martingale difference sequence error assumptions.  Through extensive simulation studies and real data analysis, we compared the performance of the proposed REVAR model with that of the standard VAR model, EVAR model, and RRVAR model. Our results consistently demonstrated the superiority of the REVAR model in terms of estimation accuracy, efficiency gains,  and forecasting performance across different scenarios and datasets. The technical details, including proofs of propositions and additional simulations, can be found in the Supplementary Materials, providing further insights into the properties of the REVAR model and its estimators.   

The REVAR model exhibits remarkable speed. For instance, fitting the model with $(d, u, p, q)=(3, 4, 1, 7)$  in Figure \ref{fig1} averaged just $0.07123$ seconds, while fitting the $(5, 10, 1, 20)$ model in Figure \ref{fig2} took approximately $0.61587$ seconds, both  with a sample size of $T = 450$. These timings were achieved using an 8-core 3.2 GHz Apple M1 processor (Table S1).  % \ref{tb.runtime}).

 For VAR models with a moderate number of variables and short lag lengths, OLS or ML estimations typically involve hundreds of parameters in coefficient and covariance matrices.   More parameters can lead to less precise estimates and larger standard errors, especially with a fixed sample size.  Applying the proposed REVAR model can parsimoniously reduce the number of model parameters, thereby improving the estimation efficiency and prediction accuracy.  Evidence suggests that increasing the VAR dimension beyond 20 in economic applications does not yield substantial gains (see Banbura et al.,2010; and Koop, 2013). However, our simulations and empirical results indicate that the REVAR model is capable of handling large datasets with up to 50 variables while maintaining robust performance. 

%Furthermore, our proposed REVAR model can be extended to a time-varying parameter model with stochastic volatility.  In recent literature, there has been a growing recognition of the importance of stochastic volatility models in forecasting, as opposed to purely time-varying coefficients (Chan and Eisenstat, 2018).  It is worth noting that any subset of components within the decomposition of  $\boldsymbol{\Sigma}_t$ in \eqref{eq:2.5} can vary over time. However,   by employing a common reducing subspace for all  $\boldsymbol{\Sigma}_t$, it becomes feasible to implement a REVAR model with stochastic volatility.    While these extensions are not straightforward, they are feasible and currently under investigation.

\vspace{-.2in}

\section*{Acknowledgments}
 \vspace{-.1in}
Thanks to the  Editor, Professor Atsushi Inoue, an Associate Editor, and anonymous referees for their insightful and constructive feedback,  which has greatly enhanced the article.  Grateful to Professors   Raja Velu and  Dennis Cook for their helpful suggestions.   %The authors report there are no competing interests to declare.

\bibliographystyle{Chicago}

\bibliography{Bibliography-MM-MC}

\begin{thebibliography}{Chicago}
%\begin{thebibliography}{12}

%\bibitem{Bosq}
%D. Bosq, {\em Nonparametric Statistics for Stochastic Processes: Estimation and Prediction}. Lecture Notes in Statistics, Vol. 110, Springer, 1998.
%\bibitem[2]{Cai}
%Z. Cai and E. Masry, {\em Nonparametric estimation of additive nonlinear ARX time series: Local linear fitting and projection}, Vol. 16, No. 4,  Econometric Theory, (2000), pp. 465-501.

%\bibitem[1]{Abraham and Ledolter}
%B. Abraham and J. Ledolter, Statistical Methods for Forecasting, (1983),  Wiley.

%\bibitem[2]{Bass and Clarke}
% F. M. Bass and D. G. Clark,  {\em  Testing distributed lag models of advertising effect}. Journal of Marketing Research, 9, (1972), pp. 298-308.

%\bibitem{Beveridge and Nelson }
%Beveridge, S.,  \&   Nelson, C.~R.  (1981).  A new approach to decomposition of economic time series into permanent and transitory components with particular attention to measurement of the `business cycle', {\it Journal of Monetary Economics}, 7, 151-174.
%
%\bibitem {Bowman}
%Bowman, A.~W., \& Azzalini, A. (1997).  {\it Applied Smoothing Techniques for Data Analysis},  London, Oxford University Press.
%
%\bibitem {Chen et al.}
% Chen, J., Kim, I., Terrel, G.~R., \&  Liu, L. (2014).  Generalised partial linear single-index mixed models for repeated measures data,  {\it Journal of Nonparametric Statistics}, 26(2), 291-303.
%
%\bibitem{Davydov}
% Davydov, Yu.~A. (1973).  Mixing conditions for Markov chains,  {\it Theory of Probability and Its Applications}, 18, 312-328.
%

%\bibitem{Efron}
%Efron, B., and Hastie, T. (2016), {\em Computer Age Statistical Inference}, Cambridge University Press, New York, NY.

%\bibitem{Frey}
%Frey, J. (2013),``Data-Driven Nonparametric Prediction Intervals", {\em Journal of Statistical Planning and Inference}, 143, 1039-1048.

\bibitem{Anderson (1951)}
Anderson, T. W. (1951), ``Estimating linear restrictions on regression coefficients for multivariate normal distributions", {\em The Annals of Mathematical Statistics}, 327-351.
%Ann. Math. Statist

\bibitem{Anderson (1999)}
Anderson, T. W. (1999), ``Asymptotic distribution of the reduced rank regression estimator under general conditions", {\em The Annals of Statistics}, 27(4), 1141-1154.
%Ann. Statist

\bibitem{Billard et al. (2023)}
Billard, L.,  Douzal-Chouakria, A. and  Samadi, S.Y. (2023), ``Exploring dynamic structures in matrix-valued time series via principal component analysis", {\em Axioms}, 12, 570.

\bibitem{Anderson (2002)}
Anderson, T. W. (2002), ``Canonical correlation analysis and reduced rank regression in
autoregressive models", {\em The Annals of Statistics}, 30(4), 1134-1154.
%Ann. Statist

% \bibitem{Bai and Ng (2002)}
% Bai, J. and Ng, S. (2002), ``Determining the number of factors in approximate factor models", {\em Econometrica}, 70(1), 191-221.

\bibitem{Banbura et al. (2010)}
Banbura, M., Giannone, D. and Reichlin, L. (2010), ``Large Bayesian vector auto regressions", {\em Journal of Applied Econometrics}, 25, 71-92.

\bibitem{Basu et al. (2019)}
Basu, S., Li, X. and Michailidis, G. (2019), ``Low rank and structured modeling of high-dimensional vector autoregressions", {\em IEEE Transactions on Signal Processing}, 67,  1207-1222.



%\blue
\bibitem{Bernardini and Cubadda (2015)}
Bernardini E. and G. Cubadda (2015), ``Macroeconomic forecasting and structural analysis through regularized reduced-rank regression", {\em International Journal of Forecasting}, 31, 682-691.

%\black 


\bibitem{Box et al. (2015)}
Box, G. E., Jenkins, G. M., Reinsel, G. C. and Ljung, G. M. (2015),  {\em``Time series analysis: forecasting and control"},  John Wiley \& Sons.

\bibitem{Box and Tiao (1977)}
Box, G. E. P. and Tiao, G. C. (1977), ``A canonical analysis of multiple time series", {\em Biometrika}, 64(2), 355-365.

\bibitem{Brune et al. (2022)}
Brune, B., Scherrer, W. and  Bura, E. (2022), ``A state-space approach to time-varying reduced-rank regression", {\em Econometric Reviews}, 41(8), 895-917.

\bibitem{Bura and Cook (2003)}
Bura, E. and Cook, R. D. (2003), ``Rank estimation in reduced-rank regression", {\em Journal of Multivariate Analysis}, 87(1), 159-176.
%J. Mult. Anal

\bibitem{Carriero et al. (2011)}
Carriero, A., Kapetanios, G. and Marcellino, M. (2011), ``Forecasting large datasets with Bayesian reduced rank multivariate models", {\em Journal of Applied Econometrics}, 26(5), 735-761.

%\blue 
\bibitem{Carriero et al. (2016)}
Carriero, A. Kapetanios, G. and M. Marcellino (2016), ``Structural analysis with multivariate autoregressive index models", {\em 
 Journal of Econometrics}, 192, 332-348.

\bibitem{Centoni and Cubadda (2015)}
Centoni M. and G. Cubadda (2015), ``Common feature analysis of economic time series: An overview and recent developments", {\em Communications for Statistical Applications and Methods}, 22, 1-20.

\bibitem{Chan et al. (2018)}
Chan, J. C. C. and Eisenstat, E. (2018), ``Bayesian Model Comparison for Time-Varying Parameter VARs with Stochastic Volatility", {\em Journal of Applied Econometrics}, 33, 509-532.
 
\bibitem{Chang et al. (2022)}
 Chang, J.,    Jiang, Q. and    Shao, X. (2022), 
``Testing the martingale difference hypothesis in high dimension",  {\em Journal of Econometrics}, In Press. 
 

 

\bibitem{Chen et al. (2013)}
Chen, K., Dong, H. and Chan, K. S. (2013), ``Reduced rank regression via adaptive nuclear norm penalization", {\em Biometrika}, 100(4), 901-920.

\bibitem{Conway (1990)}
Conway, J. (1990), {\em ``A Course in Functional Analysis"}, Spring-Verlag New York.


\bibitem{Cook and Zhang (2015a)}
Cook, R. D. and Zhang, X. (2015a), ``Foundations for envelope models and methods",  {\em Journal of the American Statistical Association}, 110(510), 599-611.

\bibitem{Cook and Zhang (2015b)}
Cook, R. D. and Zhang, X. (2015b), ``Simultaneous envelopes for multivariate linear regression", {\em Technometrics}, 57(1), 11-25.


\bibitem{Cook and Zhang (2016)}
Cook, R. D. and Zhang, X. (2016), ``Algorithms for envelope estimation", {\em Journal of Computational and Graphical Statistics}, 25(1), 284-300.

\bibitem{Cook and Zhang (2018)}
Cook, R. D. and Zhang, X. (2018), ``Fast envelope algorithms",  {\em Statistica Sinica}, 28(3), 1179-1197.

\bibitem{Cook et al. (2010)}
Cook, R. D., Li, B. and Chiaromonte, F. (2010), ``Envelope models for parsimonious and efficient multivariate linear regression (with discussion)", {\em Statistica Sinica}, 927-960.
% Statist. Sinica

\bibitem{Cook et al. (2013)}
Cook, R. D., Helland, I. S. and Su, Z. (2013), ``Envelopes and partial least squared regression", {\em Journal of the Royal Statistical Society: Series B (Statistical Methodology)}, 75(5), 851-877.
%J. R. Statist. Soc. B

\bibitem{Cook et al. (2015)}
Cook, R. D., Forzani, L. and Zhang, X. (2015), ``Envelopes and reduced-rank regression", {\em Biometrika}, 102(2), 439-456.

%\blue 
\bibitem{Cubadda  and Guardabascio (2019)}
Cubadda, G. and B. Guardabascio (2019), ``Representation, estimation and forecasting of the multivariate index-augmented autoregressive model", {\em International Journal of Forecasting}, 35, 67-79.

\bibitem{Cubadda et al. (2017)}
Cubadda G., Guardabascio B. and A. Hecq (2017), ``A vector heterogeneous autoregressive index model for realized volatility measures", {\em International Journal of Forecasting}, 33, 337-344.

\bibitem{Cubadda:Hecq(2022a)}
Cubadda, G. and Hecq, A. (2022a),  ``Reduced rank regression models in economics and finance", {\em Oxford Research Encyclopedia of Economics and Finance},  Oxford University Press. % doi: 10.1093/acre- fore/9780190625979.013.677.

\bibitem{Cubadda:Hecq(2022b)}
Cubadda, G. and Hecq, A. (2022b),  ``Dimension reduction for high dimensional vector autoregressive models", {\em Oxford Bulletin of Economics and Statistics},   84(5), 1123-1152.

\bibitem{Cubadda et al. (2019)}
Cubadda G., Hecq A. and S. Telg (2019), ``Detecting co-movements in noncausal time series, {\em Oxford Bulletin of Economics and Statistics}, 81, 697-715.
 

\bibitem{Ding and Cook (2018)}
Ding, S. and Cook, R. D. (2018), ``Matrix variate regressions and envelope models", {\em Journal of the Royal Statistical Society: Series B (Statistical Methodology)}, 80(2), 387-408.

\bibitem{Davis et al. (2016)}
Davis, R. A., Zang, P. and Zheng, T. (2016), ``Sparse vector autoregressive modeling", {\em Journal of Computational and Graphical Statistics}, 25(4), 1077-1096.

\bibitem{De Mol et al. (2008)}
De Mol, C., Giannone, D. and Reichlin, L. (2008), ``Forecasting using a large number of predictors: is Bayesian regression a valid alternative to principal components?", {\em Journal of Econometrics} 146: 318-328.

%\bibitem{Edelman et al.(1998)}
%Edelman, A., Arias, T. A. and Smith, S. T. (1998), ``The geometry of algorithms with orthogonality constraints", {\em SIAM journal on Matrix Analysis and Applications}, 20(2), 303-353.
%\blue 

\bibitem{Escanciano et al. (2006)}
Escanciano, J. C. and Velasco, C. (2006),  ``Generalized spectral tests for the martingale difference hypothesis",  {\em Journal
of Econometrics}, 134, 151–185.
%\black 

\bibitem{Forni et al. (2005)}
Forni, M., Hallin, M., Lippi, M. and Reichlin, L. (2005), ``The generalized dynamic factor model: one-sided estimation and forecasting", {\em Journal of the American Statistical Association}, 100(471), 830-840.

\bibitem{Forzani and Su (2021)}
Forzani, L. and Su, Z. (2021), ``Envelopes for elliptical multivariate linear regression", {\em Statistica Sinica}, 31, 301-332.

%\blue 
\bibitem{Franchi and Paruolo (2021)}
Franchi, M. and P. Paruolo (2011), ``A characterization of vector autoregressive processes with common cyclical features", {\em Journal of Econometrics}, 163, 105-117.
%\black 


\bibitem{Hamilton (1994)}
Hamilton, J. (1994), {\em Time series analysis}, Princeton, NJ: Princeton University Press.

% \bibitem{Hannan (1970)}
% Hannan, E. (1970), {\em Multiple time series},  New York: Wiley.

%\blue
\bibitem{Hecq et al. (2006)}
Hecq, A., Palm, F. and J. Urbain (2006), ``Common cyclical features analysis in VAR models with cointegration", {\em Journal of Econometrics}, 132, 117-141.
%\black

\bibitem{Henderson and Searle (1979)}
Henderson, H.V. and Searle, S. R. (1979), ``Vec and vech operators for matrices, with some uses in Jacobians and multivariate statistics", {\em Canadian Journal of Statistics} 7(1), 65-81.
%Can. J. Statist.

\bibitem{HerathSamadi2023a}
Herath, H. M. W. B. and  Samadi, S. Y. (2023a),  ``Partial envelope and reduced-rank partial envelope vector
autoregressive models", Preprint.

\bibitem{HerathSamadi2023b}
Herath, H. M. W. B. and  Samadi, S. Y. (2023b),  ``Scaled envelope models for multivariate time series", Preprint.

%\blue 
\bibitem{Hetland et al. (2021)}
Hetland, S., Pedersen, R. S.  and  Rahbek, A. (2021),  ``Dynamic conditional eigenvalue GARCH",  {\em Journal of Econometrics}, In press.
%\black 


\bibitem{Izenman (1975)}
Izenman, A. J. (1975), ``Reduced-rank regression for the multivariate linear model", {\em Journal of multivariate analysis}, 5(2), 248-264.
%J. Mult. Anal

\bibitem{Koop (2013)}
Koop, G. M. (2013),  ``Forecasting with medium and large Bayesian VARs", {\em Journal of Applied Econometrics}, 28(2), 177-203.

\bibitem{Lam et al. (2011)}
Lam, K. S. and Tam, L. H. (2011), ``Liquidity and asset pricing: Evidence from the Hong Kong stock market", {\em Journal of Banking \& Finance}, 35(9), 2217-2230.

\bibitem{Lam, C., & Yao, Q. (2012)}
Lam, C. and  Yao, Q. (2012), ``Factor modeling for high-dimensional time series: inference for the number of factors", {\em The Annals of Statistics}, 694-726.

%\blue 

\bibitem{Lee:Su:2020}
Lee, M. and   Su, Z. (2020),  ``A review of envelope models",  {\em International Statistical Review}, 88, 658-676.

%\black

\bibitem{Li and Zhang (2017)}
Li, L. and Zhang, X. (2017), ``Parsimonious tensor response regression", {\em Journal of the American Statistical Association}, 112(519), 1131-1146.

\bibitem{Lutkepohl (2005)}
L\"{u}tkepohl, H. (2005), {\em New introduction to multiple time series analysis}, Berlin: Springer.

%\blue
\bibitem{McCracken (2020)}
McCracken, M.W. and Ng, S. (2020), ``FRED-QD: A Quarterly Database for
Macroeconomic Research", {\em Federal Reserve Bank of St. Louis Working Paper 2020-
005}, URL https://doi.org/10.20955/wp.2020.005.
%\black

\bibitem{Negahban and Wainwright (2011)}
Negahban, S. and Wainwright, M. J. (2011), ``Estimation of (near) low-rank matrices with noise and high-dimensional scaling", {\em The Annals of Statistics}, 39(2), 1069-1097.

\bibitem{Park and Samadi (2014)}
Park, J.H. and  Samadi, S.Y. (2014),  ``Heteroscedastic modelling via the autoregressive conditional variance subspace",  {\em Canadian  Journal of  Statistics}, 42, 423–435.  

\bibitem{Park and Samadi (2020)}
Park, J.H. and  Samadi, S.Y. (2020),  ``Dimension reduction for the conditional mean and variance functions in time series", {\em Scandinavian  Journal of  Statistics}, 47, 134–155.

\bibitem{Politis et al. (1994)}
Politis, D. N. and Romano, J. P. (1994), ``The stationary bootstrap", {\em Journal of the American Statistical Association}, 89(428), 1303-1313.

\bibitem{RaO (1964)}
Rao, C. R. (1964), ``The use and interpretation of principal component analysis in applied research", {\em Sankhy\={a}: The Indian Journal of Statistics, Series A} (1961-2002), 26(4), 329-358.

\bibitem{ Raskutti et al. (2019)}
Raskutti, G., Yuan, M. and Chen, H. (2019), ``Convex regularization for high-dimensional
multi-response tensor regression", {\em The Annals of Statistics}, 47:1554-1584.

% \bibitem{Reinsel (1997)}
% Reinsel, G. C. (1997), {\em Elements of multivariate time series analysis}, 2nd Ed. New York: Springer.

\bibitem{Reinsel and Velu (1998)}
Reinsel, G. C. and Velu, R. P. (1998), {\em Multivariate Reduced-rank Regression: Theory and Applications}, New York: Springer.

%\blue

\bibitem{Reinsel et al. (2022)}
Reinsel, G. C.   Velu, R. P. and Chen, K.  (2022). {\em Multivariate reduced-rank regression: Theory, methods and applications}, New York: Springer.
%\black

\bibitem{Rekabdarkolaee et al. (2020)}
Rekabdarkolaee, H. M., Wang, Q., Naji, Z. and Fuente, M. (2020), ``New parsimonious multivariate spatial model", {\em Statistica Sinica}, 30(3), 1583-1604.

\bibitem{Samadi (2017)}
Samadi, S. Y. (2014),  ``Matrix time series analysis",  Ph.D. Dissertation, University of Georgia, Athens, GA, USA.

\bibitem{Samadi et al. (2017)}
Samadi, S.Y., Billard, L., Meshkani, M.R. and  Khodadadi, A. (2017),  ``Canonical correlation for principal components of time series",  {\em Computational
Stat.}, 32, 1191–1212.

\bibitem{Samadi and DeAlwis (2023a)}
Samadi, S. Y. and T. P. DeAlwis (2023a),  ``Fourier methods for sufficient dimension
reduction in time series",  Preprint.

\bibitem{Samadi and DeAlwis (2023b)}
Samadi, S.Y. and T. P. DeAlwis (2023b), ``Envelope matrix autoregressive models", Preprint.

\bibitem{Samadi et al. (2019)}
Samadi, S.Y., Hajebi, M. and  Farnoosh, R. (2019),  ``A semiparametric approach for modelling multivariate nonlinear time series",  {\em Canadian  Journal of Stat.}, 47, 668–687.

\bibitem{Shapiro (1986)}
Shapiro, A. (1986), ``Asymptotic theory of overparameterized structural models", {\em Journal of the American Statistical Association}, 81(393), 142-149.

\bibitem{Shojaie and Michailidis (2010)}
Shojaie, A. and Michailidis, G. (2010), ``Discovering graphical Granger causality using the truncating lasso penalty", {\em Bioinformatics}, 26(18), i517-i523.

\bibitem{Song and Bickel (2011)}
Song, S. and Bickel, P. J. (2011), ``Large vector autoregressions", {\em arXiv preprint arXiv}: 1106.3915.

\bibitem{Stock and Watson (2005)}
Stock, J. H. and  Watson, M. W. (2005), ``Implications of dynamic factor models for VAR analysis", NBER Working Paper Series w11467. %National Bureau of Economic Research, Cambridge, MA.

\bibitem{Stoica and Viberg (1996)}
Stoica, P. and Viberg, M. (1996), ``Maximum likelihood parameter and rank estimation in reduced-rank multivariate linear regressions", {\em IEEE Transactions on signal processing}, 44(12), 3069-3078.

\bibitem{Su and Cook (2011)}
Su, Z. and Cook, R. D. (2011), ``Partial envelopes for efficient estimation in multivariate linear regression", {\em Biometrika}, 98(1), 133-146.

\bibitem{Su et al. (2016)}
Su, Z., Zhu, G., Chen, X. and Yang, Y. (2016), ``Sparse envelope model: efficient estimation and response variable selection in multivariate linear regression", {\em Biometrika}, 103(3), 579-593.

\bibitem{Tiao and Tsay (1989)}
Tiao, G. C. and Tsay, R. S. (1989), ``Model specification in multivariate time series", {\em Journal of the Royal Statistical Society: Series B (Methodological)}, 51(2), 157-195.

\bibitem{Tsay (2014)}
Tsay, R. S. (2014), {\em Multivariate time series analysis: with R and financial applications}, John Wiley, Hoboken, NJ.

% \red
% \bibitem{Velu Reinsel(2013)}
% Velu R. P. and Reinsel G. C. (2013), {\em Multivariate reduced-rank regression: theory and applications}  (Vol. 136), New York: Springer-Verlag.
% \black

\bibitem{Velu et al. (1986)}
Velu, R. P., Reinsel, G. C. and Wichern, D. W. (1986), ``Reduced rank models for multiple time series", {\em Biometrika}, 73(1), 105-118.

\bibitem{Wang et al. (2022a)}
Wang, D., Zheng, Y., Lian, H. and Li, G. (2022a), ``High-dimensional vector autoregressive time series modeling via tensor decomposition", {\em Journal of the American Statistical Association}, 117:539, 1338-1356.

%\blue
\bibitem{Wang et al. (2022b)}
 Wang,  D.,   Zhang, X.,    Li, G.   and  Tsay,  R. (2022b), ``High-dimensional vector autoregression with common response and predictor factors", {\em 	arXiv:2203.15170}.

%\black 

\bibitem{Wang et al. (2021)}
Wang, D., Zheng, Y. and Li, G. (2021), ``High-dimensional low-rank tensor autoregressive time series modeling", {\em arXiv preprint arXiv}:2101.04276.

\bibitem{Wang and Ding (2018)}
Wang, L. and Ding, S. (2018), ``Vector autoregression and envelope model", {\em Stat}, 7(1), e203.
%Stat.

\bibitem{Wei, W. W. (2019)}
Wei, W. W. (2019), ``Multivariate time series analysis and applications", {\em John Wiley \& Sons}.

% \bibitem{Wei (2006)}
% Wei, W. W. S. (2006), {\em Time series analysis}, In The Oxford Handbook of Quantitative Methods in Psychology: Vol. 2.

% \bibitem{Yuan et al. (2007)}
% Yuan, M., Ekici, A., Lu, Z. and Monteiro, R. (2007), ``Dimension reduction and coefficient estimation in multivariate linear regression", {\em Journal of the Royal Statistical Society: Series B (Statistical Methodology)}, 69(3), 329-346.

%\blue
\bibitem{Zhou and Lin (2013)}
Zhou, X. C., and Lin, J. G. (2013), ``Semiparametric regression estimation for longitudinal data in models with martingale difference error's structure", {\em Statistics}, 47(3), 521-534.

\bibitem{Zhao et al. (2011)}
Zhao, Z. W., Wang, D. H., and Zhang, Y. (2011), ``Limit theory for random coefficient first-order autoregressive process under martingale difference error sequence", {\em Journal of computational and applied mathematics}, 235(8), 2515-2522.
%\black
\end{thebibliography}

 %\end{document}
 \setstretch{.9}

\end{document}